\documentstyle[12pt]{article}
\begin{document}
\title{Single- and double-beta decay Fermi-transitions
in an exactly solvable model}

\author{Jorge G. Hirsch$^1$\thanks{e-mail: hirsch@fis.cinvestav.mx},
	Peter O. Hess$^2$\thanks{e-mail: hess@roxanne.nuclecu.unam.mx}
 and
	Osvaldo Civitarese$^3$\thanks{e-mail:
civitare@venus.fisica.unlp.edu.ar}\\
{\small\it $^1$Departamento de F\'{\i}sica, Centro de Investigaci\'on
y de Estudios Avanzados del IPN,}\\
{\small\it Apdo. Postal 14-740 M\'exico 07000 D.F.}\\
 {\small\it$^2$Instituto de Ciencias Nucleares, Universidad Nacional
Aut\'onoma de M\'exico,}\\
{\small\it Apdo. Postal 70-543, M\'exico 04510 D.F.}\\
{\small\it$^3$ Departamento de F\'{\i}sica, Universidad Nacional de
La Plata, }\\ {\small\it c.c. 67 1900, La Plata, Argentina.}
}
\maketitle

\begin{abstract}
An exactly solvable model suitable for the description of single and
double-beta decay processes of the Fermi-type
is introduced.	The model is equivalent to the exact shell-model
treatment of protons and neutrons in a single j-shell.
Exact eigenvalues and eigenvectors
are compared to those corresponding to the hamiltonian in the
quasiparticle
basis (qp) and with the results of both the standard quasiparticle
random phase approximation (QRPA) and the renormalized one (RQRPA).
The  role of the scattering term of the
quasiparticle hamiltonian is analyzed.
The presence of an exact eigenstate with zero energy
is shown to be related to the
collapse of the QRPA. The RQRPA and the qp solutions do not include
this zero-energy eigenvalue in their spectra,
probably due to spurious correlations. The meaning of this result
in terms of symmetries is presented.
\end{abstract}

\noindent
PACS number(s): 21.60.Fw, 21.60.Jz, 23.40.Hc

\vskip .5cm

\section{Introduction}

In the last years the study of the Quasiparticle Random Phase
Approximation (QRPA) and its extensions, like the Renormalized
Quasiparticle Random Phase
Approximation (RQRPA), has received renewed attention.
 The goal
was to improve substantially the reliability of
the QRPA description of nuclear double beta decay transitions
and, at the same time, to enhance the predictive power
of the theory in an unambiguous way.

The predictive power of the QRPA, mostly in dealing
with the calculation of the matrix elements for
ground state to ground state two-neutrino double-beta decay
transitions ( $\beta\beta_{2\nu}$), is questionable since
these amplitudes are extremely sensitive to details of
the nuclear two-body interaction
\cite{Vog86,Eng87,Civ87,Mut89}.

The inclusion of renormalized particle-particle correlations in the QRPA
matrix amounts to a drastic suppression of the
$ \beta\beta_{2\nu}$ -matrix elements. However,
for some critical values of the model parameters; i.e: the
renormalized two-body interactions, the otherwise purely real QRPA
eigenvalue problem
becomes complex. As a consequence of it the standard properties of
the QRPA metric and
conservation rules are
severely downplayed by the appearance of strong ground-state
correlations which
jeopardize the stability of the theory.
The most notorious example of this behaviour, of the QRPA approach, is
the calculation of the $\beta\beta_{2\nu}$
decay of $^{100}Mo$ \cite{Vog86,Eng87,Civ91,Civ91b,Gri92}.

The renormalized version of the QRPA (RQRPA)\cite{Har64,Row68}, which
includes some corrections beyond the quasiboson approximation, has
been recently  reformulated \cite{Cat94} and applied to the
$\beta\beta_{2\nu}$ decay problem
\cite{Toi95}. Contrary to the QRPA, the RQRPA does not
collapse for any value of the residual two-body interaction.
Based on its properties, the RQRPA was presented as a cure
for the instabilities of the QRPA  and it was applied
to calculations of the $\beta\beta_{2\nu}$ decay of
$^{100}Mo$ \cite{Toi95}. Similar studies have been
performed in the framework of the RQRPA and with the
inclusion of proton-neutron pairing
correlations in symmetry breaking Hamiltonians. \cite{Sch96}.

In a recent paper \cite{Hir96a} we have shown that
the RQRPA violates the Ikeda sum rule and that this
violation is indeed present in many extensions of the QRPA.
The study was based on the schematic proton-neutron Lipkin model.

In a subsequent letter \cite{Hir96b} we have introduced
an exactly solvable model for the description of single- and double-beta
decay Fermi-type transitions. This model is equivalent to a single-j
shell model for protons and neutrons. The appearance of an eigenvalue
at zero energy,
in the exact spectrum, was found. Moreover, it has been shown that the
presence of this zero-energy eigenvalue should be associated to the
collapse of the QRPA. It was shown that the
RQRPA does not include this zero-energy mode in its spectrum.
It was also shown that the absence of this zero-energy
state, in the RQRPA, leads to finite but spurious
results for the transition amplitudes near the point of collapse of
the QRPA.

In the present paper we discuss the details of the exactly solvable
model of \cite{Hir96b}. The algebraic techniques needed to evaluate
matrix elements of the relevant operators, in the
SO(5) group-representation, are described in detail.
Exact eigenvalues and eigenvectors
are compared with those corresponding to the quasiparticle version of
the hamiltonian (qp) and with the ones obtained with
the QRPA and RQRPA.
The role of the correlations induced by the scattering term H$_{31}$ of
the qp-hamiltonian
and the effects on the number of quasiparticles in the ground state are
analyzed.
The presence of a zero
excitation energy state in the spectrum corresponding to the
exact solution of the model hamiltonian is discussed.
As said before it will be shown that the RQRPA and the qp
solutions, do not display the same feature, most likely due to the
presence of spurious states caused by the mixing of orders, of
the relevant interaction terms, in the expansion procedure.

The structure of the paper is the following: the model and its solutions
are presented in
Section 2, the quasiparticle version of the hamiltonian, its linear
representation in terms of pairs of unlike (proton-neutron)
quasiparticle-pairs and
its properties are introduced in Section 3. The QRPA and RQRPA
treatments of the hamiltonian are
discussed in Section 4. The matrix elements of double-beta-decay
transitions, calculated in the framework of the different approximations
introduced in the previous sections, are given
in Section 5.
Conclusions are drawn in Section 6.
The SO(5) algebra,
representations and reduced matrix elements used in
the calculations are given in detail in the Appendices A, B and C,
respectively.

\section{The model}

The model hamiltonian, which includes a single
 particle  term, a pairing term for
protons and neutrons and a schematic
 charge-dependent  residual interaction with both particle-hole
 and particle-particle channels, has been introduced in refs.
 \cite{Kuz88,Mut92,Civ94a} and it is given by

\begin{equation}
H = e_p {\cal N}_p - G_p S^\dagger_p S_p +
 e_n {\cal N}_n - G_n S^\dagger_n S_n +
2 \chi \beta^- \cdot \beta^+
	  - 2 \kappa P^- \cdot P^+ , \label{hamex}
\end{equation}

\noindent with
\begin{equation}
\begin{array}{c}
{\cal N}_i = \sum\limits_{m_i} a^\dagger_{m_i} a_{m_i} ~,\hspace{1cm}
S^\dagger_i = \sum\limits_{m_i} a^\dagger_{m_i} a^\dagger_{\bar m_i}/2 ~,
\hspace{1cm}i=p,n\\
 \beta^- = \sum\limits_{m_p = m_n} a^{\dagger}_{m_p} a_{m_n} ~,
 \hspace{1cm}
P^- = \sum\limits_{m_p = -m_n}	a^{\dagger}_{m_p} a^{\dagger}_{\bar m_n} ~;
\end{array}
\end{equation}

\noindent
$ a^{\dagger}_p = a^\dagger_{j_p m_p}$ being the particle creation
 operator and
 $a^{\dagger}_{\bar p} = (-1)^{j_p -m_p} a^{\dagger}_{j_p -m_p}$ its time
reversal. The parameters $\chi$ and $\kappa$ play the role of the
renormalization factors $g_{ph}$ and $g_{pp}$ introduced in the
literature \cite{Vog86,Eng87,Civ87,Mut89}.

 It has been shown in a series of papers
\cite{Civ94a,Civ94b,Civ95} that this
hamiltonian, when treated in the framework of the QRPA,
reproduces fairly well the results obtained with a realistic
G-Matrix constructed from the Bonn-OBEP
potential,  both for single- and double-beta decay
transitions. These results can be taken as an indication about
the correlations induced by the interactions in (1), which
are obviously specific to the relevant degrees of freedom
of the problem. In other words, if the relatively simple
schematic force (1) can approximately described the correlations
induced by a more realistic interaction it certainly means that it
is able to pick-up the bulk of the physics involved in the
transitions.

In a single-one-shell
limit, for the model space ($j_p = j_n = j$) and for monopole
($J=0$) excitations the hamiltonian (1) can be solved exactly.
In spite of the fact that the solutions obtained in this restricted
model space cannot be related to actual nuclear states,
the excitation energies,
single- and double-beta decay transition amplitudes and ground state
correlations depend on the particle-particle strength parameter
$ \kappa$ in the same way as they do in realistic
calculations with many single
particle levels and with more realistic interactions, as we shall show
later on. Physically,
the beta decay transitions between $J^{\pi}=0^+$ states
correspond to transitions of the Fermi type. However,
the study of the model and the identification of its relevant
degrees of freedom, instead of the
comparison of observables, is the main aspect of the present work.
We shall obtain the eigenstates of (1), by using different
approximations, in order to
built-up a comprehensive view about the
validity of them and their predictive power.

The hamiltonian (\ref{hamex}) can be expressed in terms of the
generators of an SO(5) algebra \cite{Par65,Hec65,Kle91}.
The Hilbert space is constructed by using the eigenstates of the
particle-number operator ${\cal N} = {\cal N}_p + {\cal N}_n$ , the
isospin $\cal T$ and its projection ${\cal T}_z = ({\cal N}_n -
{\cal N}_p)/2$. The raising and lowering isospin operators are
defined as
$\beta^\pm ={\cal T}^\pm$, where ${\cal T}^- |n\rangle = |p\rangle$.
With them we can construct the isospin scalar ${\cal T}^2 = {\frac
1 2} ({\cal T}^-{\cal T}^+ + {\cal T}^+ {\cal T}^-) + {\cal T}_z^2$
and the second order SO(5) Casimir (see Appendix A)

\begin{equation}
S^\dagger_n S_n + S^\dagger_p S_p + {\frac 1 2} P^\dagger P =
{\frac {\cal N } {\rm 4}} ( 3 - {\frac {\cal N } {\rm 2}} + 2
\Omega ) - {\frac {\cal T } {\rm 2}} ({\cal T } + 1)
\end{equation}
\noindent
with $\Omega = (2j+1)/2$.

The Hamiltonian (\ref{hamex}) can be expressed in terms of the above
mentioned operators. Hereafter we will use $G_p = G_n\equiv G$ for
simplicity. In terms of these generators the Hamiltonian (1) reads

\begin{equation}
\begin{array}{rl}
H = &[e_p +e_n - {\frac 1 3} (3 + 2\Omega - {\frac {\cal N} {\rm
2}}) (G+ 2 \kappa )] {\frac {\cal N} {\rm 2}} ~+ \\
&[e_n - e_p - 2 \chi ({\cal T}_z - 1)]{\cal T}_z ~+ \\
&[2 \chi + {\frac G 3} + {\frac 2 3} \kappa ] {\cal T} ({\cal T} +1 ) ~+ \\
&\sqrt{\frac 2 3} \Omega (4 \kappa - G )
\left [  [a^{\dagger} a^{\dagger}]^{J=0,{\cal T}=1}
[a a]^{J=0,{\cal T}=1}] \right ]^{{\cal T}=2}_{{\cal T}_z =0}
\label{hamex1} \end{array}
\end{equation}

\noindent
In writing the creation and annihilation operators
($a^\dagger ~, a$) we have omitted unnecessary subindexes
since the coupling to total angular momentum $J$ and isospin
${\cal T}$, represented as $[a^{\dagger} a^{\dagger}]^{J,{\cal T}}$,
is understood.

Hamiltonian (\ref{hamex1}) is diagonal in the ${\cal N, T, T}_z$ basis if
$G = 4 \kappa$. It can be
reduced to an isospin scalar if its parameters are selected as

\begin{equation}
e_p =e_n, \hspace{1cm}\chi=0,\hspace{1cm}G=4\kappa.
\end{equation}
If $4\kappa \neq G$ the hamiltonian (\ref{hamex}) is not diagonal
in this basis. The
hamiltonian mixes states with different isospin $T$ while its
eigenstates still have definite $N$ and $T_z$. The dynamical breaking
of the
isospin symmetry is an essential aspect of the model which is
directly related to the nuclear structure mechanism
responsible for the suppression of the
matrix elements for double-beta-decay transitions.

\subsection{The diagonal case $G = 4\kappa$}

The solution of (1) in the basis $\mid {\cal N,T,T}_z >$,
in the case $G = 4\kappa$, gives a state, the isobaric analog state (IAS)
at the energy

\begin{equation}
\begin{array}{ll}
E_{IAS} &= E({\cal N, T, T}_z={\cal T}-1 ) - E({\cal N, T, T}_z
={\cal T})  \\ &= e_p - e_n + 4 \chi ({\cal T} -1)
\end{array}
\end{equation}

Considering a double Fermi-transition, the energy available
for the decay is given by

\begin{equation}
\begin{array}{ll}
E_{\beta\beta} &= E({\cal N, T, T}_z) - E({\cal N, T', T}_z' ) \\
&= 2 (e_n - e_p) + 8 \chi + G (2 {\cal T} -1) ~,
\hspace{.5cm}\hbox{if}~{\cal T} \ge 2~. \end{array}
\end{equation}
\noindent
where ${\cal T}'_z= {\cal T}_z-2$. The above expression shows
clearly the role of the particle-hole
strength parameter
$\chi$. It determines the excitation energy of the IAS, which
depends not only upon the proton-neutron energy shift due to the
nuclear Coulomb field but also upon $\chi$ and $\cal{T}$.
The same dependence is shown by the $Q_{\rm{value}}$
 (eq.(7)). In analogy
with the situation found in realistic calculations
its value can be determined by a fit
to the experimental value of the IAS energy (or to the Gamow-Teller
Giant Resonance for the case of $J^{\pi} = 1^+$
spin-isospin-dependent excitations).

The $\beta$ decay operators for single
Fermi transitions, ${\cal T}^\pm$, do not change
the total isospin neither the total particle number of the
state upon which they act. Only the isospin
projection of the state is changed in steps of one unit,
namely:

\begin{equation}
{\cal T}^\pm ~|{\cal N T T}_z \rangle
= \sqrt{( {\cal T}\pm {\cal T}_z +1) ({\cal T}\mp {\cal T}_z ) }
	~|{\cal N T T}_z \pm 1 \rangle
\end{equation}

\subsection{The spectrum}

For the numerical examples we have selected
$N_n > N_p $ and a large value of $j$ to
simulate the realistic situation found in medium-  and heavy-mass
nuclei. To perform the calculations we have adopted the following
two sets of parameters:

\begin{equation}
\begin{array}{llll}
\hbox{set I} &j = 9/2, & N = 10, & 0 \le T_z \le 4,\\
&e_p = 0.96 MeV, & e_n = 0.0 MeV, \\
& G_p = G_n = 0.4 MeV, &\chi= 0~ \hbox{or}~0.04 MeV~,
&0 \le \kappa \le 0.2\\~~\\
\hbox{and} & & & \\
\hbox{set II} &j = 19/2, & N = 20, & 1 \le T_z \le 5,\\
&e_p = 0.69 MeV, & e_n = 0.0 MeV, \\
& G_p = G_n = 0.2 MeV, &\chi=0 ~\hbox{or}~ 0.025 MeV~,
&0 \le \kappa \le 0.1
\end{array}
\end{equation}

The dependence of the spectrum and transition matrix elements on
the parameters $\chi$ and $\kappa$ is analyzed in the following
paragraphs.

 {\bf Fig. 1}

The complete set of $0^+$ states, belonging to different isotopes,
is shown in Fig. 1a and Fig. 2a, for $G= 4\kappa$ and $\chi = 0.$,
as a function of the number of protons (Z).
The states are labeled by the isospin quantum numbers $(T,T_z)$.
Ground states are shown by thicker lines. As shown in these
figures the structure of the mass parabola is qualitatively
reproduced.

The upper insert, case a) of each figure, shows the full spectrum
corresponding to
$\chi = 0$. The lower one, case b),
shows the results corresponding to $\chi = 0.05 MeV$ (Fig 1.b)
and  $\chi = 0.025 MeV$ in (Fig 2.b).
Obviously the particle-hole channel of the residual
interaction stretches the spectra of all isotopes. As mentioned above,
it increases the energy of the IAS.

Beta decay transitions of the Fermi  type,
mediated by the action of the operator $\beta^- =  t^-$,
are allowed between states belonging to the same isospin multiplet.
The
energy of each member of a given multiplet increases
linearly with $Z$.

In this example
the  $0^+$ states belonging to each odd-odd-mass nuclei ( N-1, Z+1, A)
are the IAS constructed from
the $0^+$ states of the even-even-mass nuclei with
(N, Z, A) nucleons. Thus, Fermi transitions between them are allowed.

Since the isospin of the ground state of each of
the even-even-mass nuclei
differs, for different isotopes,
Fermi-double-beta-decay transitions connecting them are forbidden
in this diagonal limit $G = 4 \kappa$.

\subsection{Exact solutions}

The Hamiltonian (\ref{hamex}) has a ${\cal T}=2$
tensorial component
which mixes states with different isospin, while particle number
and isospin projection remain as good quantum numbers.	The
diagonalization of (1) is performed
in the basis of states described in the Appendix B. The corresponding
reduced matrix elements are given in the Appendix C.
The eigenstates are written as:

\begin{equation}
|{\cal N T}_z \alpha \rangle = \sum\limits_{\cal T} {\cal
C}^\alpha_{{\cal N T T}_z} |{\cal N T T}_z \rangle
\end{equation}

{\bf Fig. 3}

\bigskip

The energy of the ground-state ($0_{g.s}^+$) and of the first-excited
state ($0_1^+$), as a function of the ratio $4 \kappa /G$
for the set of parameters $j=9/2,~ {\cal N}_n = 6, ~{\cal N}_p =
4,~ \chi= 0$ are shown if Fig. 3a. The results of Fig.3b have
been obtained with the set of parameters given by
$j=19/2,~{\cal N}_n = 12, ~{\cal N}_p = 8$ and $\chi=0$.

The most characteristic feature of the results is the barely
avoided crossing of levels, due to the repulsive
nature of the effective residual
interaction between them. Although a complete level-crossing
is not obtained in this model,
in the neighbourhood of the value
$4 \kappa /G \approx 1 $
a major structural change in the wave
functions will develope. In the case of a complete crossing of levels
the ground and the first excited state will interchange
their quantum numbers thus given raise to a permanently deformed
(in the sense of the isospin dominance) situation.

This behaviour is by no means a surprise
since it is similar to that found in
pairing plus quadrupole systems \cite{Kis63}. In this case, if the
quadrupole-quadrupole interaction is strong enough, the system
becomes permanently deformed, in the sense of the angular momentum
and spatial rotations.	The analogy between this and the
present case ( isospin degree of freedom) can be drawn
from the study of
\cite{Aga68,Sch68} where
the "pairing plus monopole" model, which is a
two-level model exactly solvable using the SO(5) algebra, was used
to analyzed the spherical and the deformed
regime of the solutions of the multipole-multipole interaction.

\bigskip
{\bf Figs. 4,5}

\bigskip

The full-thin line of Fig. 4a (4b) represents the excitation
energy	$E_{exc}$ of the lowest $0^+$ state belonging to the
double-odd-mass
nucleus (${\cal N}_n=7, {\cal N}_p = 3$) with respect to the parent
even-even-mass nucleus
(${\cal N}_n=8, {\cal N}_p =2$) as a function of the ratio
 $4 \kappa / G$ for $j=9/2$ and $\chi
= 0$~ ($0.04$).
It is clear that when $4 \kappa / G \approx 1.6  (1.8)$ attractive
proton-neutron
correlation dominates over  proton-proton and neutron-neutron
pairing correlations and the  excitation energy goes to zero.
Similar results are depicted in Figs. 5a and 5b, corresponding to
the excitation
energy	$E_{exc}$ of the lowest $0^+$ state in the odd-odd
mass
nucleus (${\cal N}_n=13, {\cal N}_p= 7$), also measured
from the ground state of the parent
even-even nucleus with
(${\cal N}_n=14, {\cal N}_p=6$) , for $j=19/2$ and $\chi=0$  ($0.025$).
In the case of Fig.5
the excitation energy goes to zero  when $4 \kappa / G \approx 1.3 $.

The vanishing of the energy of the first excited state,
and the subsequent inversion of levels (or negative excitation energies)
would indicate that the double-odd nucleus becomes more bound
than their even-even neighbours, contradicting the main
evidence for the dominance of like-nucleons pairing
in medium- and heavy-mass nuclei.
It would also completely suppress
the double beta decay because the single beta decay from each
"side" of the double-odd nucleus would be allowed.

These result simply emphasizes the fact that the
Hamiltonian (\ref{hamex}) will not be the adequate one when
attractive proton-neutron interactions are too large.
In a realistic situation, obviously, the true Hamiltonian
includes other degrees of freedom, like
quadrupole-quadrupole interactions,  and permanent deformations
of the single-particle mean-field can also be present.
These additional degrees of freedom will prevent the complete
crossing of levels which, of course, is not observed. However,
in many cases the experimentally observed energy-shift of double-odd-mass
nuclei, respect to their double-even-mass neighbours
is very small. This finding reinforces the notion of an underlying
dynamical-symmetry-restoration-effect.

\section{The Hamiltonian in the quasiparticle (qp) basis}

By performing the transformation of the particle creation and
annihilation operators of the Hamiltonian (1) to the quasiparticle
representation
\cite{Row70};i.e. by using the Bogolyubov transformations for
protons and neutrons,
we have obtained
the Hamiltonian
\begin{equation}
\begin{array}{rl}
H = &(\epsilon_p -\lambda_p) N_p + (\epsilon_n - \lambda_n) N_n +
\lambda_1 A^\dagger  A + \lambda_2 ( A^\dagger A^\dagger + A A )  \\
&-\lambda_3 ( A^\dagger B + B^\dagger A) -
\lambda_4 (A^\dagger B^\dagger + BA) + \lambda_5 B^\dagger B +
\lambda_6 (B^\dagger B^\dagger + B B) \label{hambcs}
\end{array}
\end{equation}

\noindent
where
$\epsilon_p, \epsilon_n$ are the quasiparticle
energies, $\lambda_p, \lambda_n$ the chemical potentials and

\begin{equation}
\begin{array}{l}
A^\dagger   = \left [ \alpha^{\dagger}_p \otimes
\alpha^{\dagger}_n \right ]^{J=0}_{M=0} ,\hspace{.5cm}
B^\dagger   = \left [ \alpha^{\dagger}_p \otimes
\alpha_{\bar n} \right ]^{J=0}_{M=0} ,\hspace{.5cm}
N_i = \sum\limits_{m_i}  \alpha^{\dagger}_{im_i}
\alpha_{im_i}\hspace{.3cm} \hbox{i=p,n} \\
\lambda_1 = 4\Omega \left [\chi (u_p^2 v_n^2 + v_p^2 u_n^2) -
\kappa (u_p^2 u_n^2 + v_p^2 v_n^2 ) \right ] ~,\hspace{.5cm}
\lambda_2 = 4\Omega ( \chi + \kappa ) u_p v_p u_n v_n ~,\\
\lambda_3 = 4\Omega ( \chi + \kappa ) u_n v_n (u_p^2-v_p^2)~,\hspace{.5cm}
\lambda_4 = 4\Omega ( \chi + \kappa ) u_p v_p (u_n^2-v_n^2) ~,\\
\lambda_5 = 4\Omega \left [\chi (u_p^2 u_n^2 + v_p^2 v_n^2) -
\kappa (u_p^2 v_n^2 + v_p^2 u_n^2 )\right ] ~,\hspace{.5cm}
\lambda_6 = - \lambda_2 ~.
\end{array}
\end{equation}

 The operators $A^\dagger$ ($A$), which  create (annihilate)
a pair of unlike (proton and neutron)-quasiparticles, together
with their counterparts for pairs of
identical quasiparticles and $B, B^\dagger, N_p, N_n$ are the generators
of the SO(5) algebra \cite{Par65}.

The quasiparticle energies
\begin{equation}
\epsilon = G \Omega /2~. \label{eqp}
\end{equation}
and the occupation probabilities
\begin{equation}
v_p^2 = {\frac {{\cal N}_{\rm p}} {\rm 2j+1} } ~~,\hspace{1cm}
v_n^2 = {\frac {{\cal N}_{\rm n}} {\rm 2j+1} }~.
\end{equation}
are determined from the gap equations and particle-number conservation
condition \cite{Row70}.
The occupation probabilities can also be defined in terms
of the single-particle and quasiparticle energy, namely:

\begin{equation}
v_i^2 = {\frac 1 2} ( 1 - {\frac {e_i - G v_i^2 - \lambda_i}
{\epsilon_i}} ) ~, \hspace{1cm} i = p,n
\end{equation}

\noindent

 From this equation and from Eq. (\ref{eqp}) the
chemical potentials can be expressed as

\begin{equation}
\lambda_i = e_i - {\frac {G \Omega} 2} + G v_i^2 ~,
\hspace{1cm} i=p,n
\end{equation}

The excitation energy, $E_{exc}^\lambda $, of a state
 $|0_\lambda \rangle$
belonging to the spectrum of a double-odd mass nucleus,
with ${\cal N}_p +1$ protons
and ${\cal N}_n -1$
neutrons, respect to the ground state of the even-even
neighbour with ${\cal N}_p, {\cal N}_n$, can be easily calculated
if blocking is considered,
i.e. when $v_p, v_n$ are calculated for the even-even and odd-odd
nuclei separately. These excitation energies are given by

\begin{equation}
E_{exc}^\lambda =  E(\lambda,{\cal N}_p +1,{\cal N}_n -1) -
E(g.s.,{\cal N}_n,{\cal N}_p) + \lambda_p -\lambda_n	\label{eexc}
\end{equation}

In the following we shall always refer to Eq. (\ref{eexc}) as a
suitable approximation for the excitation energies.
In the present calculation we have selected $e_p - e_n$ is
such a way that
\begin{equation}
\lambda_p - \lambda_n~=~e_p -e_n - {\frac G 2} ({\cal N}_n -{\cal N}_p)
{\frac {\Omega -1} \Omega} ~=~ 0~,
\end{equation}

\noindent
which implies

\begin{equation}
e_p -e_n~ =~ {\frac G 2} ({\cal N}_n -{\cal N}_p)
{\frac {\Omega -1} \Omega} ~.
\end{equation}

Alternatively, one can compute the occupation amplitudes
$v_p, v_n$ always for the
even-even
nucleus, without including blocking. The
effect of blocking on the unperturbed excitation
energies, with
$\kappa=\chi=0$, can be ignored if the
single-particle energy difference between protons and neutrons
is modified to the value

\begin{equation}
e_p -e_n~ =~ {\frac G 2} ({\cal N}_n -{\cal N}_p -1)
{\frac {\Omega -1} \Omega} ~.
\end{equation}

The linearized version of the Hamiltonian  (\ref{hambcs}) is
obtained by
keeping only the first line of Eq. (\ref{hambcs}). This is equivalent
to neglect terms
proportional to $B$ and $B^\dagger$ (the so-called scattering terms).
The solutions of this truncated Hamiltonian have been discussed
in a previous paper \cite{Hir96a}.

Finding the eigenvalues and eigenvectors  of Hamiltonian
(\ref{hambcs}) requires the use of the same algebraic techniques
involved in
solving the original Hamiltonian. However, the	complexity of the
problem increases severely, due to the fact that neither the
quasiparticle number or the quasiparticle isospin projection
(or equivalently the number of proton and neutron quasiparticles)
are good quantum numbers. It implies that the dimension of the basis
will increase by two orders of magnitude. Additional reduced matrix
elements would then be needed
to diagonalize the Hamiltonian (\ref{hambcs}). The analytic
expressions of these matrix elements are given in Appendix C.

There is a remaining symmetry in  Hamiltonian (\ref{hambcs}), since
states with
even number of proton-	and neutron-quasiparticles
are not connected with states having an odd number
of them. Due to this fact
it is possible to diagonalize separately these two cases.

Particle number is not a good quantum number, obviously, because
it is broken spontaneously by the Bogolyubov
transformation. Thus, zero-quasiparticle states belonging
the even-even-mass nucleus have good average number of protons and
neutrons, the condition used to determine $v_p, v_n$, while states
with a non-vanishing number
of quasiparticles show
strong fluctuations in the particle number. Fluctuations in the
particle number can induce, naturally,
important effects on the observables. Moreover, the
admixture of several quasiparticle-configurations in a given state,
induced by residual particle-particle interactions, can also strongly
influence the behaviour of the	observables. An example of this
effect is
given in \cite{Hir96a}, concerning the violation of the
Ikeda Sum Rule produced by
large values of the particle-particle strength $\kappa$.

The spectrum of the qp-Hamiltonian (\ref{hambcs}) is shown in
Figures 4 and 5.
The curves shown by small-dotted-lines, in Figs. 4.a), 4.b),
 5.a), 5.b),
display the dependence of the
excitation energy for the  qp-hamiltonian (\ref{hambcs}) upon
the ratio $\kappa /G$. The results of this qp-approximation
closely follow the exact ones
up to the point where they become negatives ($4 \kappa /G \approx 1.4~
-~1.8$ in the different cases). From this point on
they vanish, rather than taking negative values, instead.
The excitation energies for the linearized hamiltonian $H_{22}+H_{04}$
are shown as thick lines in these figures. We can see that the
linearized hamiltonian is able to reproduce qualitatively the
behaviour of the full-qp one, but in general it overestimates the
values of the excitation energies.

As it is mentioned above, the results shown in Figures 4 and 5
have been obtained both with the complete qp-hamiltonian and
 with the truncated hamiltonian which includes
only the product of pair-creation and annihilation-
operators. In \cite{Hir96a} the relevance of the scattering terms in
(\ref{hambcs}) was pointed out. From the present results
it can be seen that the inclusion of these terms is indeed
important if one looks after a better description
of the qp-excitation energies, up to
the point where the exact excitation energies become negative.
For
larger values of $\kappa$ even the eigenstates of the complete
hamiltonian fail to describe negative excitation energies. This is
a clear indication that other
effects can play an important role, like, i.e; effects associated
to the appearance of spurious
states. This can be quantitatively illustrated by the following.
There are four exact eigenstates for
$j=19/2, ~{\cal N}_n = 13, ~{\cal N}_p =7$, as can be seen in Fig. 2.a),
while the spectrum of the qp-hamiltonian (\ref{hambcs})
has 220 eigenstates.
It is
well known that states with ${\cal N}_n = 14 \pm N_n, ~{\cal N}_p = 6
\pm N_p$, where $N_p$ and $N_n$ are the number of quasiparticle protons
and neutrons,
respectively, are mixed with two-(p-n)-quasiparticle states
in the
odd-odd nucleus and provide a large number of states belonging to
other nuclei. When $4 \kappa /G \ll 1$ the spurious states remain
largely un-mixed with the lower energy two-qp state. But when
$4\kappa/G \approx 1$ the mixing becomes important.
This fact up-grades the relevance of particle-number violation
effects in dealing with this case.

The full qp-treatment represents the best possible extension of
the quasiboson
approximation, without performing a
particle-number projection, in a single-j shell. It goes beyond any
second extended RPA \cite{Mar90} and it includes explicitly all
number of proton-
and neutron-quasiparticles ($N_p$ and $N_n$) in the eigenstates.

To analize the effects associated to the number of
quasiparticles in the ground state of double-even nuclei,
and particularly the effects associated to the number of
quasiprotons,
we have calculated the average number of quasi-protons
using the expression
\begin{equation}
\langle 0_\lambda | N_p |0_\lambda \rangle =  \sum\limits_{N T T_z}
| C^\lambda_{N T T_z} |^2  (N/2 + T_z)~.
\end{equation}

A similar expression holds for the average
 neutron-quasiparticle-number.

{\bf Fig 6, 7}

In Figs. 6.a) and 6.b) the average number of proton-quasiparticles in the
ground state of the
even-even nucleus with ${\cal N}_p=6,~{\cal N}_n=14$ is shown as a
function of $4 \kappa/G$, for $j=19/2,
~\chi=0 $ and $0.04$. Figs. 7.a) and 7b) show the number of
proton-quasiparticles for
${\cal N}_p=8,~{\cal N}_n = 12, ~j = 19/2, ~\chi =0$ and $0.025$. The
dashed-lines represent the results corresponding to the full
qp-hamiltonian case while the large dots refer to the linearized $H_{22}
+ H_{04}$ version of it.
The difference between both approximations is evident. Using the
linearized hamiltonian
the states are composed only by proton-neutron-quasiparticle pairs
\cite{Hir96a}, while the presence
of the scattering terms introduces also like-(p-p and n-n)-quasiparticle
pairs. The presence of these pairs, which for
$4\kappa /G \approx 1$ play a crucial role, increases
notably the number of
quasiparticles and yields excitation energies closer to the exact ones.

The average quasiparticle number shows a saturation in the full-qp
case for $4\kappa /G \approx 1.8$.
At this value of the residual pn-interaction
the ground state is far-away for the qp vacuum, and has a structure
which can be described as a full quasiparticle shell. Notice that
at this point the
exact and full-qp excitation energies depart from each other. A state
with four proton
and four neutron quasiparticles has very large number-fluctuations.
Spurious states
become strongly mixed with physical states. In this
way the resulting excitation
energies average to zero, a limit which
differs from the exact value, which is negative.

The differences in the average qp number between full-qp and linearized
approaches are larger in Figs.6.a) and 6.b) than in Figs.7.a) and
7.b). This result is a consequence of the dependence of
some of the effective couplings of
Eq. \ref{hambcs}, i.e:	$\lambda_3$ $\approx$ $u_p^2-v_p^2$ and
$\lambda_4$
$\approx$ $u_n^2 -v_n^2$, on the number of particles.
As the number of particles approaches the saturation value $\Omega =
{\frac {2j+1} 2}$ the effective couplings
$\lambda_3$ and $\lambda_4$ will vanish. Thus, the full-qp and
linearized solutions yield similar results.

\section{QRPA and RQRPA}

The QRPA hamiltonian $H_{QRPA}$ can be obtained from the linearized
version of Eq. (\ref{hambcs}), by keeping only the bilinear-terms
in the pair-creation and pair-annihilation operators.
The pair-creation and pair-annihilation operators, $A^{\dagger}$ and A,
are, of course defined by coupled pairs of fermions. The commutation
relations between these pseudo-boson operators include number-like
quasiparticle operators in addition to unity. By
taking the limit $(2j+1) \rightarrow \infty$ \cite{Hir96a}
these extra terms vanish and the commutation relations between
pairs of fermions can be treated like exact commutation relations
between bosons. This is the well-known quasi-boson approximation and
the QRPA hamiltonian is the leading order hamiltonian which
satisfies the quasi-boson approximation. If the pair-operators
are replaced by  quasi-bosons, the resulting Hamiltonian is given by

\begin{equation}
H_{QRPA} = (2 \epsilon + \lambda_1) ~b^\dagger b ~+
\lambda_2 \{ ~b^\dagger b^\dagger ~+ ~b b \}.
\end{equation}

As said above, the quasi-bosons $b^\dagger$ and $ b$
fulfill exactly the commutation relation $[b, b^\dagger ] = 1$.

At this point we can refer to pair of fermions ( $A^{\dagger}$) or to
quasi-bosons ($b^{\dagger}$) without lost of generality, since we
have not introduced a particular representation for the pair of
fermions to boson mapping.

The QRPA states are generated by the action of the one-phonon operator
$O^\dagger_{QRPA} = X A^\dagger - Y A$ on the correlated QRPA
vacuum $|0\rangle$.
The quasi-boson approximation assumes that $\langle0| [A,A^\dagger
]|0\rangle = 1$ and it leads to the normalization condition
$X^2 - Y^2 = 1$.
The QRPA matrix is just a $ 2 \times 2$ one, with sub-matrices
$ {\cal A_{QRPA}} =  2 \epsilon + \lambda_1$ and ${\cal B_{QRPA}} = 2
\lambda_2$.
The corresponding eigenvalue is given by
$ E_{QRPA} = [(2 \epsilon + \lambda_1)^2 - 4
\lambda_2^2]^{1/2}$. It becomes purely imaginary if $2 \lambda_2
 > 2 \epsilon + \lambda_1$.

For this limit the backward-going amplitudes of the QRPA
phonon-operator become dominant, thus invalidating the
underlying assumption about the smallness of the
quasi-boson vacuum-amplitudes.
The QRPA excitation energies, obtained with the above introduced
Hamiltonian are shown in Figs. 4 and 5.
It can be seen that in the four cases displayed in these figures
the collapse of the QRPA values occurs near
the point where the exact excitation energies become negative.
This is a very important result because it means that
that the QRPA description of
the dynamics given by the hamiltonian (\ref{hamex}) is able to reproduce
exact results. At this point one can naturally ask the obvious
question about the nature of the mechanism
which produces such a collapse. The fact that the QRPA approximation
is sensitive to it, together with the fact that the
same behaviour is shown by the exact solution, reinforces the idea
about the onset of correlations which terminate the regime of
validity of the pair-dominant picture.
In order to identify such correlations we have calculated the
expectation value of the number of quasi-fermions and bosons
on the	QRPA ground state.

The average number of proton
quasiparticles in the QRPA ground state,
which in this case coincides with the average
boson number, is given
by\footnote{Notice that there is a factor 2
missprinted in Eq. (19) of ref.\cite{Hir96a}.}
\begin{equation}
\langle 0 | N_p | 0 \rangle = Y^2 ~.
\end{equation}

Figs. 6.a), 6.b), 7.a) and 7.b) show the results corresponding to these
occupation numbers. The QRPA results extend up to
the value $4 \kappa
/ G \approx 1$, where the QRPA collapses. The sudden increase of the
average quasiparticle
number near the collapse of the QRPA is a clear evidence
about a change in the structure of
the QRPA ground state.

In the renormalized QRPA the structure of the ground state is
included explicitly \cite{Row68}, in the form
\begin{equation}
|0\rangle = {\cal N} e^S |BCS\rangle~, \hspace{.5cm}S = {\frac {c
A^\dagger A^\dagger}
{2 \langle 0| [A,A^\dagger ]|0\rangle } }\hspace{.3cm}.
\end{equation}
where the quasi-boson approximation, at commutator's level, is not
enforced explicitly. The renormalization procedure consists of
retaining approximately the number of quasiparticle-like-terms of the
commutators keeping them as a parameter to be determined, namely,
by defining the
RQRPA one-phonon state as
\begin{equation}
 O^\dagger_{RQRPA} |0\rangle = \left [{\cal X} A^\dagger - {\cal Y} A \right ]
/ \langle 0| [A,A^\dagger ]|0\rangle^{1/2} |0\rangle \end{equation}

\noindent
and enforcing the condition $ O_{RQRPA} |0\rangle = 0$, which
leads to the estimate
$c = {\cal Y/X}$ for the parameter entering in the definition of the
correlated vacuum. After some algebra
it is possible to show that $\langle0| [A,A^\dagger ]|0\rangle	\equiv D
 = 1 - {\frac {2 {\cal Y}^2 D} {2j+1} } $ \cite{Cat94,Toi95}, and that
\begin{equation}
 D = \left [ 1 + {\frac {2 {\cal Y}^2 } {2j+1} } \right ]^{-1}.
\end{equation}

 The RQRPA submatrices are ${\cal A_{RQRPA}} =	2 \epsilon + \lambda_1 D$
 and ${\cal B_{RQRPA}} = 2 \lambda_2 D$. Since $0 \leq D \leq 1$, the
 presence of $D$ multiplying both $\lambda_1$ and $\lambda_2$  produces
 the
reduction of the residual interaction  which is needed to avoid
the collapse of
the QRPA equations \cite{Toi95}.
Due to this fact the RQRPA energy $E_{RQRPA}$ is always real. Its
value can be obtained by solving simultaneously the non-linear equations
for $E_{RQRPA}, {\cal X,Y}$ and $D$, which in the general case will
include all possible values of the multipolarity $J$ \cite{Toi95}.

 RQRPA	excitation energies are shown in Figs. 4 and 5.
These results strongly resemble those of Fig. 1 of ref. \cite{Toi95}
and Fig. 2 of ref. \cite{Civ91b}. The
results corresponding to the QRPA (of ref.\cite{Civ91b}) and to the
RQRPA (of ref. \cite{Toi95}) are quite similar to those shown
in Figures 4 and 5. However, the main finding of the present
calculations is that the exact
excitation energies are closer to the QRPA energies, rather than to
the renormalized ones, instead.
In exact calculations including the spin degrees of freedom a phase
transition was found at the point where the QRPA collapses \cite{Eng96},
thus reinforcing the present results.

The average number of quasiparticles in the RQRPA vacuum is given by
\begin{equation}
\langle 0 | N_p | 0 \rangle = {\cal Y}^2 ~,
\end{equation}

\noindent
and it is shown in Fig. 6.a), 6.b), 7.a), and 7.b).
It is fairly obvious, from these results, that the RQRPA
ground state correlations double in all the cases those of the complete
solutions
of the linearized hamiltonian. This is clearly an overestimation, and it
is probably one of the most
notorious difficulties confronting the use of the RQRPA.

It allows too
much ground state
correlations, and with them the particle number fluctuations are
introducing spurious
states which can dominate the low energy structure for
large values of $\kappa$.

Near "collapse" the average number of quasiparticles given by the QRPA
and the RQRPA are comparable.
For the case of the QRPA the increase of the ground-state-correlations
is determined by the change in the sign of the backward-going
matrix relative to the forward-going one near collapse. From
there on the QRPA
cannot produce any physically acceptable result since one of
the underlying conditions of the approximation, i.e: the
positive definite character of either linear combination of the
forward- and backward-going blocks of the QRPA matrix will not be
fulfilled. This collapse is prevented in the RQRPA, by the
use of the renormalization of the matrix elements, but the drawback
of the approximation is the
contribution coming from spurious states, which ought to be removed.
Moreover, there are several other reasons to cast doubts on the
consistency of the RQRPA. Among them, the mixing-up of orders
in the wave functions, of the RQRPA phonons, is not accompanied
by the the enlargement of the hamiltonian, to accommodate other
correlations, like: i.e: the exchange terms of the QRPA matrix.
If one performs such a calculation, by including exchange terms, the
resulting values of the QRPA matrix terms are also "renormalized",
but this effect will depend upon the configurations. Also, the
point of collapse is shifted to higher values of the coupling constant
$\kappa$ but the effect is tipically of the order  $1/\Omega$, as
compared to leading order terms. If terms others than unity are
introduced in the commutators, then the hamiltonian has to be enlarged
to account for the $AB$ sort of terms of the initial hamiltonian,
see eq.(11),
because they will contribute at the same order as the added
number-type of terms introduced by the RQRPA procedure. Thus,
the RQRPA procedure should be accompanied by a renormalization
of the transition operators and/or by the inclusion of
scattering terms also in these operators. At this level,
by going beyond the leading order QRPA approximation, more
terms have to be added to the diagrams which represent the
transition amplitudes.
It has been done for a pure seniority model in \cite{Duk96}.
This approach, for correlations
between pairs of like-quasiparticles, is already cumbersome
and it introduces an unmanageable number of contributions,
both to the QRPA matrix as well as to the transition operators.
For unlike-pairs of quasiparticles the situation can be even worse,
since the complete algebra, which supports the expansions, cannot be
defined
in a subspace where scattering terms are replaced by c-numbers.
More details about these aspects will be presented in a forthcoming
publication.

\section{Double beta decay}

 In this section we shall briefly discuss some of the consequences
of the previously presented approaches on the calculation
of nuclear double-beta decay observables. In the following
we shall focus our attention on the two-neutrino mode of the
nuclear double-beta decay, since the matrix elements governing
this decay mode are more sensitive to nuclear structure effects
than the ones of the neutrinoless mode. As said in the
 introduction we shall
consider only double-Fermi transitions.
The nuclear matrix elements of the two-neutrino
double-beta-decay  $M_{2\nu}$ can be written as:
\begin{equation}
 M_{2\nu} ~=~ \sum_\lambda {\frac {\langle0_f|\beta^-| 0_\lambda \rangle
\langle0_\lambda|\beta^- |0_i\rangle } {E_\lambda -E_i + \Delta} }
~,\label{m2nu}
\end{equation}

\noindent
where $ |0_i\rangle , |0_\lambda $ and $\rangle |0_f\rangle $ represent
the initial,
intermediate and final nuclear states participant of the virtual
transitions entering in the allowed second-order weak-processes.
The energies of the initial and intermediate states are $E_i$ and
$E_\lambda$, respectively. The energy released by the decay is
represented by the quantity $\Delta$. For the present calculations
we have selected the value of $\Delta = 0.5~MeV$, which is of the
order of magnitude of the empirical values used in realistic calculations.
The results for the matrix elements $M_{2\nu}$, obtained with the
exact wave functions are
shown, as a function of the ratio $4 \kappa/ G$,
in Figs. 8.a) and  8.b). These results have been obtained
with the following set of parameters:
$j=9/2, ({\cal N}_p=2,~{\cal N}_n=8)\rightarrow ({\cal N}_p=4,~{\cal
N}_n=6)$ and $\chi =
0 $ and $0.04 MeV$, respectively. The values shown in
Figs. 9.a) and	9.b) correspond to
$j=19/2, ({\cal N}_p=6,~{\cal N}_n=14)\rightarrow ({\cal
N}_p=8,~{\cal N}_n=12)$ and $\chi = 0 $ and $0.025 MeV$.

{\bf Fig. 8,9}

For all cases the exact value of
 $M_{2\nu}$ vanishes at the point $4 \kappa / G = 1$.
As mentioned
above, this cancellation
appears in the model due to the fact that for this value of $\kappa$ the
isospin-symmetry is recovered
and the ground states of the initial
and final nuclei
belong to different isospin multiplets, as it can be seen also
from the results shown in Figs. 1 and 2.

A similar mechanism, in the context of a solvable model
possessing a SO(8) algebra including spin
and isospin degrees of freedom
was used a decade ago to show that the
cancellation of the
$M_{2\nu}$ matrix elements for certain values of the particle-particle
residual interaction was
not an artifact of the QRPA description \cite{Eng87}.

The results corresponding to the matrix elements $M_{2\nu}$,
calculated with the different approximations discussed in the
text are shown in Figs.8 and 9, as a function of the coupling
constant $\kappa$.
The values  of $M_{2\nu}$ are very similar to those
found in realistic
calculations \cite{Vog86,Civ87,Mut89,Toi95}, including its strong
suppression for values of the coupling constant $\kappa$
near the value which produces the collapse of the QRPA description.
Distinctively, the RQRPA results extends to values
of $\kappa$ passing the "critical" value. However, the validity of this
result can be questioned because, as we have shown above,
the RQRPA missed the vanishing of the excitation energy.
The $M_{2\nu}$ matrix elements, evaluated
with the complete qp-hamiltonian
(\ref{hambcs}), is quite similar to that of
the RQRPA up to point where it vanishes.
 From this point-on the results of both the full-qp
and the RQRPA approximations  are different. Both matrix elements
change their sign at a
value of $\kappa$ which is larger than the one corresponding
to the change of the sign of the matrix elements calculated with the
exact wave function. The fact that the RQRPA results and the ones of
the qp-approximation are similar, although these models differ
drastically in the correlations which they actually include, suggest
that a kind of balance is established between terms which are
responsible for ground state correlations and those which produce
the breaking of coherence in the wave functions. Obviously this
mechanism must be related to the presence of scattering terms
in the commutators as well as in the Hamiltonian.

\section{Conclusions}

An exactly solvable model for the description of single- and double-beta-
decay-processes of the Fermi-type was introduced.
The model is equivalent to a complete shell model treatment in a
single-j shell for the adopted hamiltonian.
It reproduces the main features of the results obtained in
realistic calculations, with
many shell and effective
residual interaction, like those used in the literature to describe the
microscopic structure of the nuclei involved in double
beta decay processes.

We have constructed the exact spectrum of the Hamiltonian
and discussed its properties. The results concerning the
energy of the states belonging to the exact solution of the
model show that, in spite of its very schematic structure,
the hamiltonian is able to qualitative reproduce the nuclear mass
parabola. The sequence of levels of the exact
solution shows
that the ground-state and the first-excited state,
of the spectrum of double-even nuclei, approach
a band-crossing situation for a critical value of the
strength associated to attractive particle-particle interactions.
At the crossing these states interchange their quantum numbers.
This behaviour is connected with the
description of "shape" transitions in similar theories, where the
order parameter is clearly associated with multipole deformations
in r-space. In the present model the "deformation" mechanism is
related with the breaking of the isospin symmetry and the space-rotation
correspond to a rotation in isospin-space which preserves the
third-component of the isospin.

We have compared the exact values of the excitation energy and of the
double-beta-decay matrix elements, for double-Fermi transitions,
with those obtained by using the solutions of the
approximate qp-hamiltonian, its linearized version and both the QRPA and
RQRPA ones.

It was shown that the collapse of the QRPA correlates with
the presence of
an exact-eigenvalue at zero energy. The structure of the RQRPA solutions
has been
discussed and it was found that though finite they are not free from
spurious contributions. The role of scattering-terms was discussed and
they were shown to be relevant in getting excitation energies closer
to the exact values. However they are not enough
to generate the correlations which are needed to produce
the band-crossing, or negative excitation energies, as it was found
in the exact solution  for large values of the coupling constant
$\kappa$.

In order to correlate the break-up of the QRPA approximation
with the onset of strong fluctuations in the particle number
we have calculated the average number of quasiparticles in the
different approximations discussed in the text.

It was shown that the solutions of the complete qp-hamiltonian
display a strong change in
the structure of the ground state when the particle-particle strength
increases.
The qp-content of the ground state varies
from a nearly zero-value to an almost full qp-occupancy.
The particle
number fluctuations
associated with states with a large number of quasiparticles were
mentioned as a possible source of spurious states.

Double beta decay amplitudes were evaluated in
the different formalisms. Their
similitudes and differences were pointed out.

As a conclusion the need of additional
work, to clarify the meaning of the different approximations
possed by the RQRPA, was pointed out.

\section{Acknowledgments}

Partial support of the Conacyt of Mexico and the CONICET of
Argentina is acknowledged. One of the authors (O.C) gratefully
acknowledges a grant of the J. S. Guggenheim Memorial
Foundation.

\newpage

\appendix{\large\bf Appendix A: the SO(5) algebra}

\bigskip

Following \cite{Par65} we introduce the operators
\begin{equation}
\begin{array}{l}
A^\dagger (0)  \equiv \left [ \alpha^{\dagger}_p \otimes
\alpha^{\dagger}_n \right ]^{J=0}_{M=0} = A^\dagger,\\
A^\dagger (1)  \equiv {\frac 1 {\sqrt{2}} }\left [
\alpha^{\dagger}_n \otimes
\alpha^{\dagger}_n \right ]^{J=0}_{M=0} ,\hspace{.5cm}
A^\dagger (-1)	\equiv	{\frac 1 {\sqrt{2}} } \left [
\alpha^{\dagger}_p \otimes \alpha^{\dagger}_p \right ]^{J=0}_{M=0} \\
B^\dagger   = \left [ \alpha^{\dagger}_p \otimes
\alpha_{\bar n} \right ]^{J=0}_{M=0} ,\hspace{.5cm}
T^- = - \sqrt{ 2\Omega} B^\dagger,  \label{gen1}
\end{array}
\end{equation}
\noindent
which together with their hermitian conjugates and with the number
and isospin operators
\begin{equation}
 N = N_p + N_n, \hspace{.3cm}  T_z = {\frac {N_p - N_n} 2}, \hspace{.3cm}
N_i = \sum\limits_{m_i}  \alpha^{\dagger}_{im_i}
\alpha_{im_i}\hspace{.3cm} \hbox{i=p,n}  \label{gen2}
\end{equation}
are the ten generators of the SO(5) group.

The hermitian conjugates of the pair-creation operators transform, under
isospin-reversal like
\begin{equation}
\tilde A(M) \equiv (-1)^{M} A(-M).
\end{equation}

Their commutation relations are more easily expressed defining the
new operators

\begin{equation}
\begin{array}{ll}
H_1 = {\frac N 2} - \Omega ,& H_2 = T_z ,\\
E_{11} = \sqrt{\Omega} A^\dagger(1), &E_{-1-1} = \sqrt{\Omega} A(1), \\
E_{1-1} = -\sqrt{\Omega} A^\dagger(-1), &E_{-11} = -\sqrt{\Omega} A(-1),\\
E_{10} = \sqrt{\Omega} A^\dagger(0), &E_{-10} = \sqrt{\Omega} A(0), \\
E_{01} = {\frac 1{\sqrt{2}} } T^+ , &E_{0-1} = {\frac 1{\sqrt{2}} } T^- .
\end{array}
\end{equation}

The operators $E_{\alpha \beta}$ are raising and lowering
 operators. When operating on an eigenstate of the weight operators
$H_1$ and $H_2$ they increase or decrease the
eigenvalues of one or both by one unit.
 Their commutation relations are

\begin{equation}
\begin{array}{l}
[H_1,H_2] = 0, \hspace{.5cm} [H_1, E_{\alpha \beta} ] = \alpha
E_{\alpha \beta} , \hspace{.5cm} [H_2, E_{\alpha \beta}] = \beta
E_{\alpha \beta}, \\

[E_{\alpha \beta}, E_{-\alpha -\beta}] = \alpha H_1 + \beta H_2 , \\

[E_{\alpha \beta}, E_{\alpha' \beta'}] =
\left \{
\begin{array}{l}
\pm ~E_{\alpha + \alpha' \beta + \beta'} ~~\hbox{if~} \alpha + \alpha'
\hbox{~and~} \beta + \beta' = 0,1,-1 \\
0 ~~~\hbox{otherwise}
\end{array} \right.
\end{array}
\end{equation}

More explicitly
\begin{equation}
\begin{array}{lll}
[E_{11}, E_{-10}] = E_{01}, & [E_{11}, E_{0-1}] = - E_{10},
& [E_{10}, E_{-11}] = - E_{01},\\

[E_{10}, E_{-1-1}] =  E_{0-1}, & [E_{10}, E_{01}] = - E_{11},
& [ E_{10}, E_{0-1}] = E_{1-1},\\

[E_{1-1}, E_{-10}] = - E_{0-1}, &  [E_{1-1}, E_{01}] = E_{10}
\end{array}
\end{equation}

\noindent
and by hermitian conjugation of the above
commutators one obtains

\begin{equation}
E_{\alpha \beta}^\dagger = E_{-\alpha -\beta}.
\end{equation}

\bigskip

\appendix{\large\bf Appendix B: SO(5) representations}

\bigskip
The highest weights of the operators $H_1,H_2$ define the
irreducible representations (irrep) of the SO(5) algebra.
For the present
case we want the irrep which contains the state with zero
quasiparticles as well as the state completely filled with
quasi-proton and
quasi-neutrons. The maximum number ($N_{max}$) of quasiparticles
allowed by the Pauli principle is $2 \Omega$, thus adding
quasi-protons and quasi-neutrons  one obtains $N_{max} = 4\Omega$.
This is state with
the highest weight and it belongs to the irrep defined
by $(H_1 = \Omega , H_2 =
0 )$ or $N = 4\Omega,T=T_z =0$. Acting with the generators
(\ref{gen1}) on this state it is possible to generate the set of
all the states with even number of quasiparticles. This
subspace suffices for all the calculations described
in this work. For this reason we have adopted the
irrep $(H_1 = \Omega , H_2 = 0 )$.

In general it is necessary to specify four quantum numbers to
completely define a state in a given irrep. But for the present case
it turns out that the states can be defined by the quantum
numbers $N, T, T_z$.

In the following we will construct explicitly the states $|N T
T_z=T\rangle$; others states with $T_z \ne T$ are obtained by acting
with the isospin lowering operator $T^-$ on them.
The states of this basis are defined by

\begin{equation}
\begin{array}{l}
|N ~T=T_z\rangle = N(a,b) (O_{00})^b(O_+)^a |N=4\Omega ~T=T_z=0\rangle,
~~~\hbox{where}\\
O_+ = E_{-11}, \hspace{.5cm} O_{00} = 2 E_{-11}E_{-1-1} + E_{-10}
E_{-10},\\
a = T_z = T, \hspace{.5cm} b = \Omega - {\frac T 2} - {\frac N 4},\\
N(a,b) = 2^b
\left [ {\frac
{ (2 \Omega + 1 - 2b)! (\Omega -a-b)! (2a+1)! (a+b)! }
{ (2 \Omega +1)! (\Omega -b)! (a!)^2 b! (2a+2b+1)! }
} \right ] ^{1/2}
\end{array}
\end{equation}

\newpage

\appendix{\large\bf Appendix C: SO(5) reduced matrix elements}
\bigskip

To diagonalize hamiltonian (\ref{hambcs})  in the $( N, T, T_z)$
basis, or hamiltonian (\ref{hamex})  in the ${\cal N, T, T}_z$ basis,
requires the use of the Wigner-Eckart theorem
\begin{equation}
\langle N' T' T_z' | O^{n t t_z} | N T T_z \rangle =
 ( T T_z, t t_z | T' T_z' )
\langle N' T' || O^{n t} || N T \rangle
\end{equation}

\noindent
where in the right hand side the symbol $(..,..|..)$ represents
a Clebsch-Gordan coefficient and $\langle..||[..]||..\rangle$
is a reduced matrix
element.
Explicit expressions for the reduced matrix elements are given below.
The
difference in the number of creation and annihilation operators in
the tensor $O$ is represented by $n$ and in order to obtain
non-zero matrix elements it must be equal
to $N'-N$.

We have used of the Wigner-Eckart theorem, the
commutation relations given in the Appendix A and the explicit form of
the states with $T=T_z$ shown in the Appendix B
to calculate the reduced matrix elements of the different operators
which are relevant in our problem. Some of these SO(5)-reduced
matrix elements are listed here.
Additional matrix elements can be deduced from them by using

\begin{equation}
\langle N T || (O^{n t})^\dagger || N' T' \rangle =  \sqrt{\frac {2T'+1}
{2T+1} } \langle N'~T' || O^{n t} || N T \rangle ~.
\end{equation}

The relevant reduced matrix elements are

\begin{eqnarray*}
\lefteqn{\langle N ~T+2||[A^\dagger \tilde A]^{t=2}||N T\rangle ~=~
{\frac {-1} {2\Omega}} }\\
&&
\left [ {\frac
{ (2\Omega - T - N/2) (T+N /2+3) (-T+N/2)
( 2\Omega +T-N/2 +3)(T+1)(T+2)}
{(2T+3)(2T+5)} } \right ]^{1/2} ~,
\end{eqnarray*}

$$
\langle N T||[A^\dagger \tilde A]^{t=0}||N T\rangle =
{\frac {1} {2 \sqrt{3}\Omega}}
\left [  (2\Omega - N/2 + 3) N /2 -T(T+1) \right ] ~,\\
$$

\begin{eqnarray*}
\lefteqn{\langle N T||[A^\dagger \tilde A]^{t=2}||N T\rangle ~=~
{\frac {1} {\sqrt{6} (T T,2 0|T T)}}
\left [ \langle N T=T_z|A^\dagger(1) A(1) |N T=T_z\rangle + \right. }
\\
&&
\left. \langle N~ T=T_z|A^\dagger(-1) A(-1) |N~ T=T_z\rangle -
2 \langle N~ T=T_z|A^\dagger(0) A(0) |N~ T=T_z\rangle \right ] ~,
\end{eqnarray*}

\noindent where

\begin{eqnarray*}
\lefteqn{\langle N~ T=T_z | A^\dagger(1) A(1)|N~ T=T_z\rangle ~=~ }\\
&& {\frac {1} {\Omega}}
\left [ -\Omega +T+N/2 +
{\frac {(2\Omega - T - N/2)(T+N /2+3)(T+1)} {2 (2T+3)} } \right ]~,
\end{eqnarray*}

\begin{eqnarray*}
\lefteqn{\langle N~ T=T_z | A^\dagger(-1) A(-1)|N~ T=T_z\rangle ~=~ }\\
&& {\frac {1} {\Omega}}
\left [
{\frac {(2\Omega - T - N/2 + 3)(-T+N /2)(T+1)} {2 (2T+3)} } \right ]~,
\end{eqnarray*}

\begin{eqnarray*}
\lefteqn{\langle N~ T=T_z | A^\dagger(0) A(0)|N~ T=T_z\rangle ~=~ {\frac {1}
{\Omega}} \left [ -\Omega + N/2 +  \right. }\\
&& \left. {\frac {(2\Omega - T - N/2)(T+N /2+3)\Omega}
{(2\Omega +T-N/2+1)(-T+N/2+2)} }
\langle N+4 ~T=T_z |A^\dagger(0) A(0)|N+4 ~T=T_z\rangle  \right ] ~.\\
\end{eqnarray*}

The largest value that $N$ can take is $4\Omega - 2T$. In this
case
\begin{eqnarray*}
\langle N=4\Omega-2T ~T=T_z |A^\dagger(0) A(0)|N=4\Omega -2T ~T=T_z\rangle = 1
- T/\Omega
\end{eqnarray*}

The above reduced matrix elements are enough to deal with
hamiltonian (\ref{hamex}), which conserves particle number.
Working with hamiltonian (\ref{hambcs}) requires many other
reduced matrix elements, like the following matrix elements

\begin{eqnarray*}
\lefteqn{\langle N+4~ T||[A^\dagger A^\dagger]^{t=0}||N T\rangle ~=~
{\frac {-1} {2 \sqrt{3} \Omega}}  }\\
& \left [
(T+N/2+3) (-T+N/2+2) (2\Omega -T-N/2)(2\Omega +T-N/2+1)
 \right ]^{1/2} ~,
\end{eqnarray*}

\begin{eqnarray*}
\lefteqn{\langle N+4~ T-2||[A^\dagger A^\dagger]^{t=2}||N T\rangle ~=~ {\frac 1
{2\Omega}}  }\\
& \left [ {\frac
{ (-T+N/2+4) (-T+N/2+2) (T-1) T (2\Omega +T-N/2-1)(2\Omega +T-N/2+1)}
{(2T-1) (2T-3) } } \right ]^{1/2} ~,
\end{eqnarray*}

\begin{eqnarray*}
\lefteqn{\langle N+4~ T+2||[A^\dagger A^\dagger]^{t=2}||N T\rangle ~=~ {\frac 1
{2\Omega}}  }\\
& \left [ {\frac
{ (T+N/2+3) (T+N/2+5) (T+1) (T+2) (2\Omega -T-N/2)(2\Omega -T-N/2-2)}
{(2T+3) (2T+5) } } \right ]^{1/2} ~,
\end{eqnarray*}

\begin{eqnarray*}
\langle N+2~ T-1|| A^\dagger ||N T\rangle ~=~
 - \left [ {\frac { T (-T+N/2+2) (2\Omega +T-N/2+1)}
{2 \Omega (2T-1)  } } \right ]^{1/2} ~,
\end{eqnarray*}

\begin{eqnarray*}
\langle N+2~ T+1|| A^\dagger ||N T\rangle ~=~
  \left [ {\frac { (T+1) (T+N/2+3) (2\Omega -T-N/2)}
{2 \Omega (2T+3)  } } \right ]^{1/2} ~.
\end{eqnarray*}

These matrix elements, together with those associated with the isospin
raising and lowering operators
\begin{eqnarray*}
\lefteqn{T^+ = -\sqrt{2\Omega} B ~,\hspace{1cm}T^- = -\sqrt{2\Omega}
B^\dagger ~,} \\
&\langle N ~T ~T_z+1 | B | N ~T ~T_z\rangle = - [(T+T_z+1)(T-T_z)]^{1/2} /
\sqrt{2\Omega} ~, \\
&\langle N ~T ~T_z-1 | B^\dagger | N ~T ~T_z\rangle = -
[(T-T_z+1)(T+T_z)]^{1/2} / \sqrt{2\Omega} ~,
\end{eqnarray*}

\noindent
are all the elements which are needed to diagonalize the hamiltonian
(\ref{hambcs}) and to calculate the matrix elements of the
transition operators.

\newpage

\centerline{\bf Figure Captions}

\bigskip
Figure 1.a (1,b):  $0^+$ states of different isotopes are shown
for $j=9/2$, $4\kappa /G =1$ and $\chi = 0. (0.05) ~MeV$, in an energy
vs. $Z$ plot.
States are labeled by $(T,T_z)$.
The lowest energy state of each nucleus is
shown by a thick-line.

\bigskip
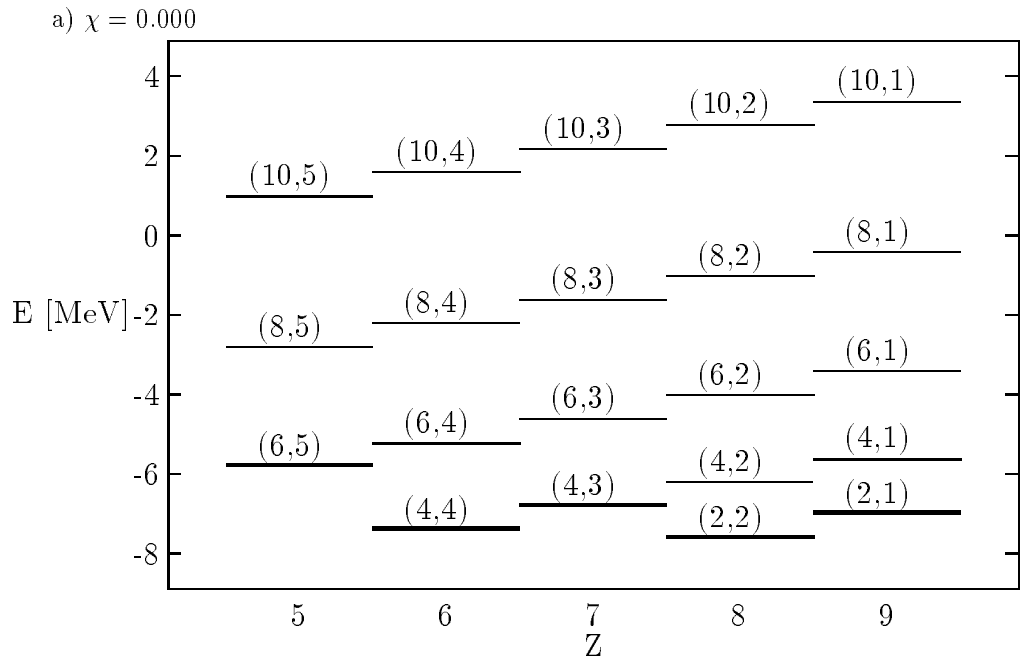
Figure 2.a (2.b): The same as Fig 1 for $j=19/2, ~4\kappa /G =1$
and $\chi = 0. ~(0.025) MeV$.

\bigskip
Figure 3.a): Energy of the ground state $0_{gs}^+$ (full line)
and first
excited
state $0_1^+$ (dotted line), as a function of the ratio
$4 \kappa /G$, for $j=9/2,~ {\cal
N}_n = 6, ~{\cal N}_p = 4,~ \chi = 0$. Figure. 3.b) shows
the same quantities for
$j=19/2,~{\cal N}_n = 12, ~{\cal N}_p = 8$.

\bigskip
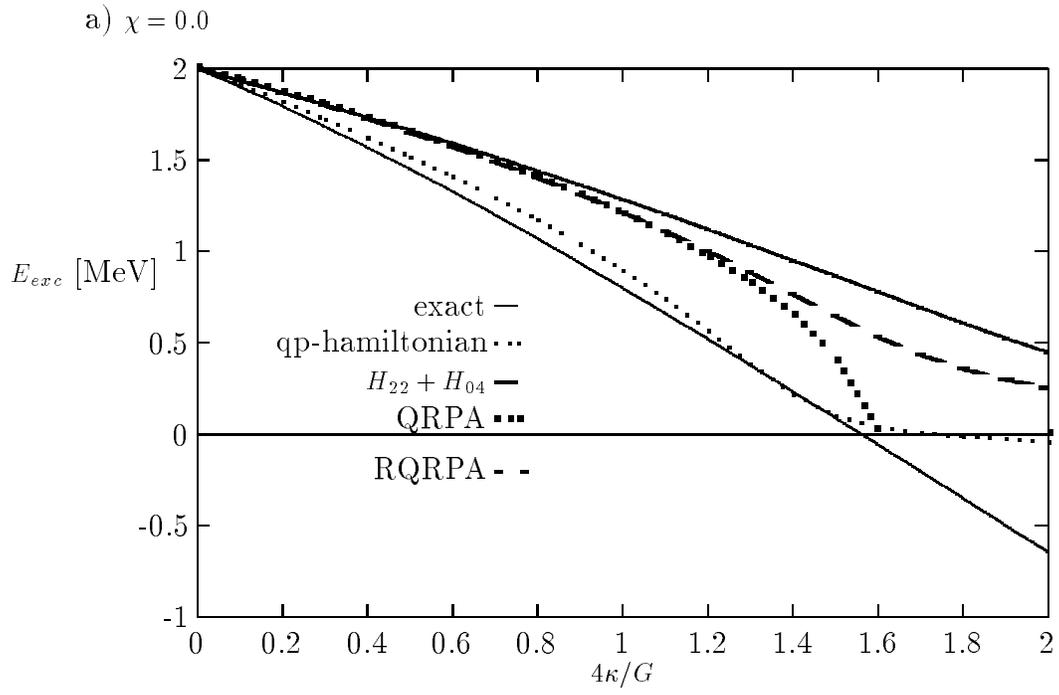
Figure 4.a (4.b): Excitation energy  $E_{exc}$ of the lowest $0^+$
state in the odd-odd
intermediate nucleus (${\cal N}_n=7, {\cal N}_p = 3$) with
respect to the
parent even-even nucleus
(${\cal N}_n=8, {\cal N}_p =2$) against $4 \kappa / G$ for
 $j=9/2,~ \chi
= 0$~ ($0.04$).
Exact results are shown as thin-full-lines while those of the
qp-hamiltonian are shown as small-dotted-lines. Results
corresponding to the linearized qp-hamiltonian
are shown as full-thick-lines and the results obtained with
the QRPA and RQRPA methods as large-dotted- and dashed-lines,
respectively.

\bigskip
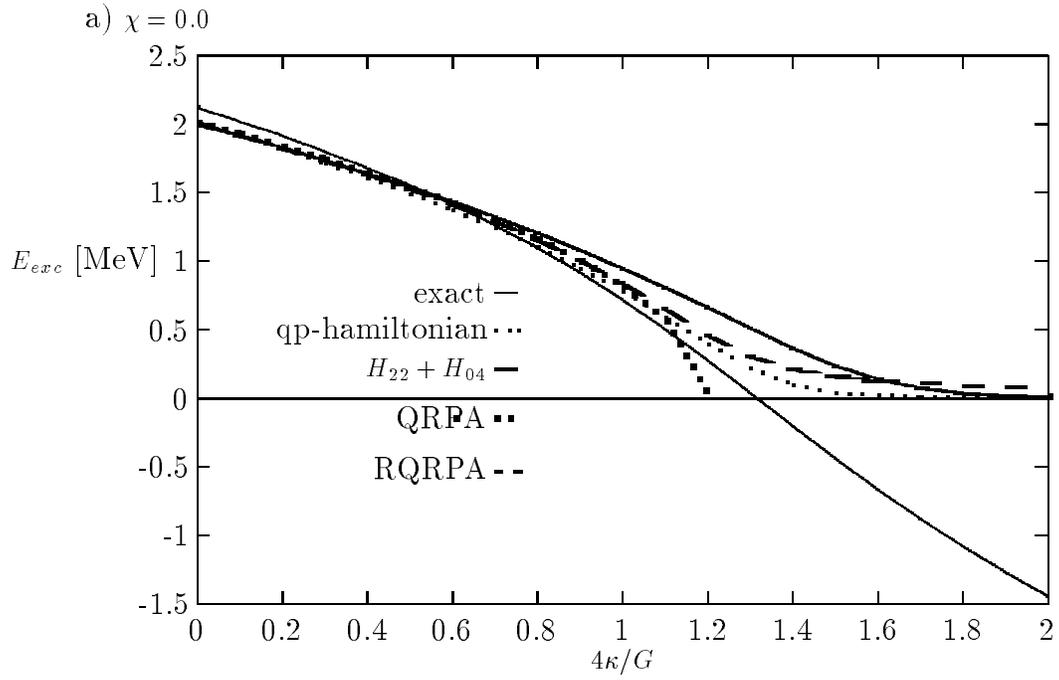
Figure 5.a (5b): The same as Figure 4.a) but for the excitation
energy	$E_{exc}$ of the lowest $0^+$ state in the odd-odd
intermediate
nucleus (${\cal N}_n=13, {\cal N}_p= 7$) with respect to the parent
even-even nucleus
(${\cal N}_n=14, {\cal N}_p=6$) , for $j=19/2,~\chi=0$	($0.025$)

\bigskip
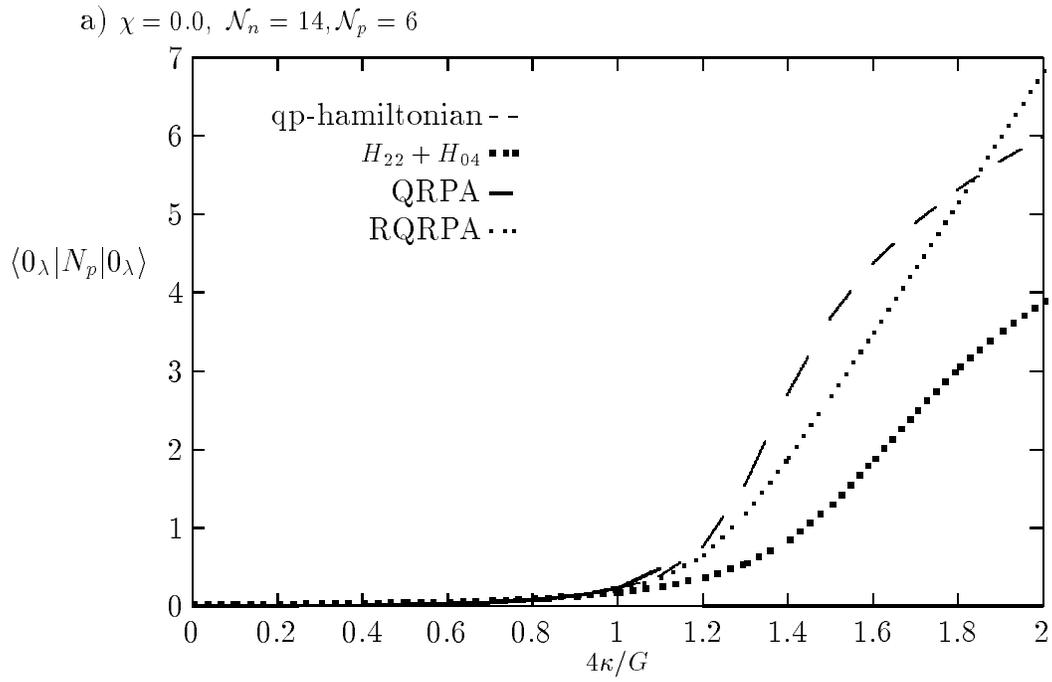
Figure 6.a (6.b):  Average number of proton-quasiparticles
 in the ground
state of the
even-even nucleus with ${\cal N}_p=6,~{\cal N}_n=14$  as
 function of $4
\kappa/G$, for $j=19/2, ~\chi=0 ~(0.025) MeV$.
Results corresponding to the qp-hamiltonian are shown as
dashed-lines. The ones corresponding to the linearized
 qp-hamiltonian are shown  as large-dotted lines and those of the
QRPA and RQRPA methods as full-lines and small-dotted-lines,
respectively.

\bigskip
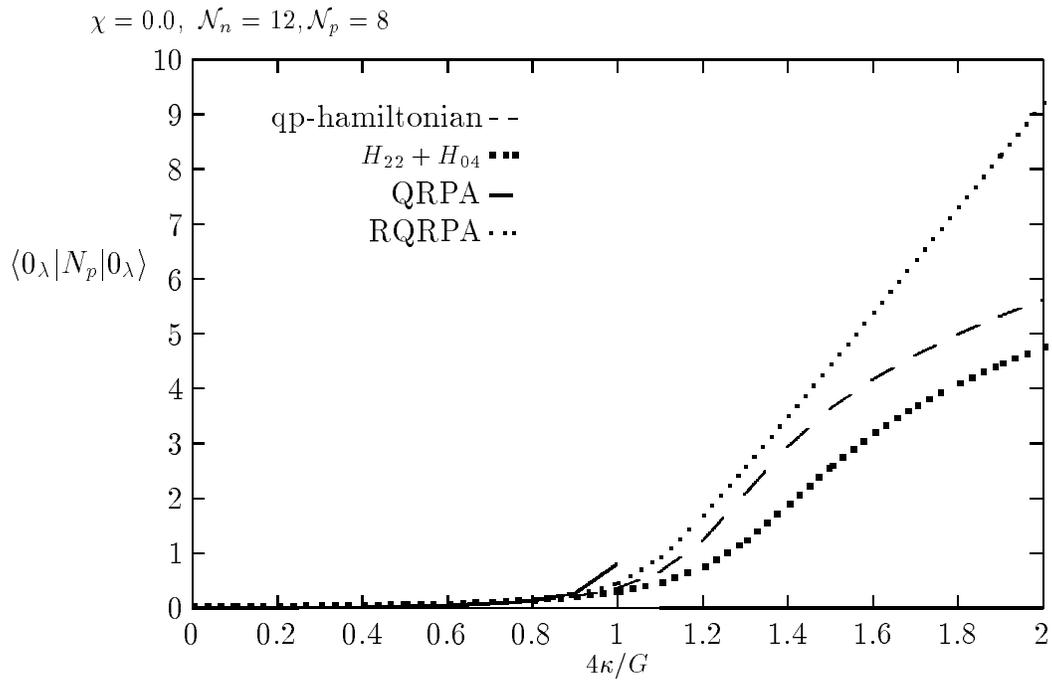
Figure 7.a (7b): The same as Figure 6 for the number of
proton-quasiparticles for
${\cal N}_p=8,~{\cal N}_n = 12, ~j = 19/2, ~\chi =0 ~(0.025) MeV.$

\bigskip
Figure. 8a (8b): Matrix elements  $M_{2\nu}$, for the
double-Fermi two-neutrino double-beta decay mode, as
functions of the ratio $4 \kappa/ G$
for $j=9/2, ({\cal N}_p=2,~{\cal N}_n=8)\rightarrow ({\cal
N}_p=4,~{\cal N}_n=6)$ and $\chi = 0
~(0.04) MeV$. Exact results are indicated by thin-full-lines.
The results obtained with the
qp-hamiltonian are shown as small-dotted-lines and the results of the
QRPA and RQRPA methods as dashed-lines and large-dotted-lines,
respectively.

\bigskip

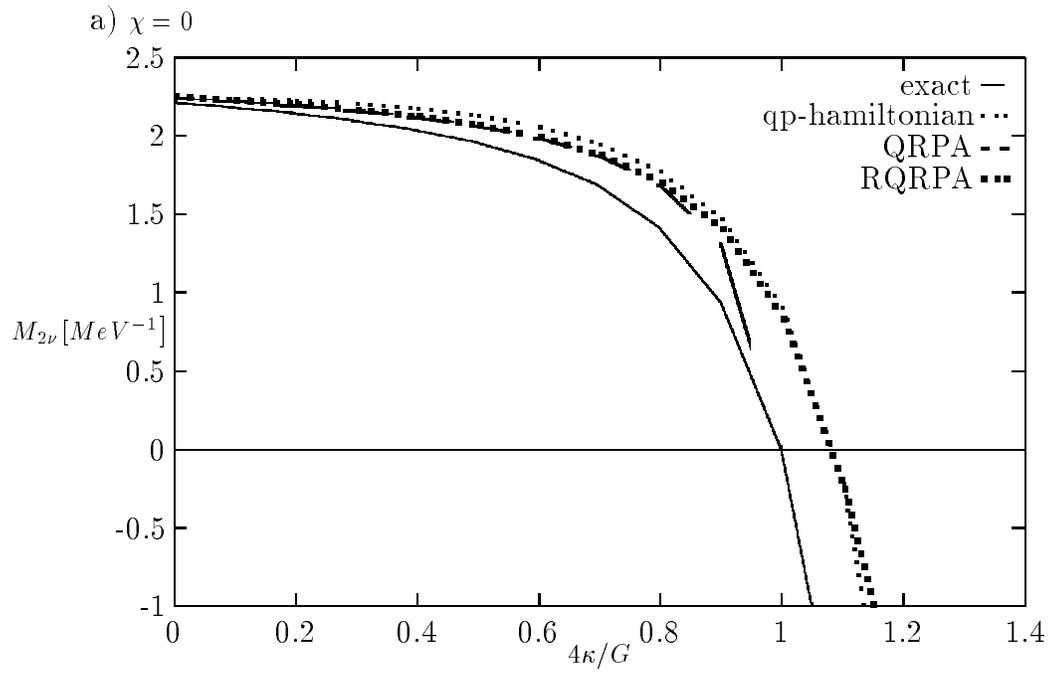
Figure 9.a (9.b): The same as Figure 8), i.e: the matrix elements
$M_{2\nu}$,
for $j=19/2, ({\cal N}_p=6,~{\cal N}_n=14)\rightarrow ({\cal
N}_p=8,~{\cal N}_n=12)$ and $\chi = 0 ~(0.025) MeV$


\bigskip
\begin{thebibliography}{Hir94b}
\bibitem{Vog86} P. Vogel and M. R. Zirnbauer, Phys. Rev. Lett. {\bf 57}
(1986) 3148.
\bibitem{Eng87} J. Engel, P. Vogel and M. R. Zirnbauer, Phys. Rev {\bf
C37} (1988) 731.
\bibitem{Civ87} O. Civitarese, A. Faessler and T. Tomoda, Phys.
Lett. {\bf B194} (1987) 11.
\bibitem{Mut89} K. Muto, E. Bender and H. V. Klapdor, Z. Phys. {\bf
A334} (1989)
\bibitem{Civ91}  O. Civitarese, A. Faessler, J. Suhonen and X. R. Wu,
Nucl. Phys. {\bf A524} (1991) 404.
\bibitem{Civ91b}  O. Civitarese, A. Faessler, J. Suhonen and X. R. Wu,
 J. Phys. G: Nucl. Part. Phys. {\bf 17} (1991) 943.
\bibitem{Gri92} A. Griffiths and P. Vogel,  Phys. Rev. {\bf C 46}
(1992) 181.
\bibitem{Har64} K. Hara, Progr. Theor. Phys. {\bf 32} (1964) 88;
K. Ikeda, T. Udagawa and Y. Yamaura, Prog. Theo. Phys. {\bf 33}  (1965) 22.
\bibitem{Row68} D. J. Rowe, Phys. Rev. {\bf 175} (1968) 1283; Rev. Mod.
Phys. {\bf 40} (1968) 153; J. C. Parick and D. J. Rowe, Phys. Rev. {\bf
175} (1968) 1293; D. J. Rowe, Nucl. Phys. {\bf A 107} (1968) 99.
\bibitem{Cat94} F. Catara, N. Dinh Dang, M. Sambataro, Nucl. Phys. {\bf
A 579} (1994) 1.
\bibitem{Toi95} J. Toivanen and J. Suhonen, Phys. Rev. Lett. {\bf 75}
(1995) 410.
\bibitem{Sch96} J. Schwieger, F.Simkovic and Amand Faessler, Nucl. Phys.
{\bf A 600} (1996) 179.
\bibitem{Hir96a} J. G. Hirsch,	P. O. Hess and O. Civitarese, Phys.
Rev. {\bf C 54} (1996) 1976.
\bibitem{Hir96b} J. G. Hirsch,	P. O. Hess and O. Civitarese, Phys. Lett.
{\bf B} in press.
\bibitem{Kuz88} V. A. Kuz'min and V. G. Soloviev, Nucl. Phys. {\bf A
486} (1988) 118.
\bibitem{Mut92} K. Muto, E. Bender, T. Oda and H. V.
Klapdor-Kleingrothaus. Z. Phys. A - Hadrons and Nuclei {\bf 341}
(1992) 407.
\bibitem{Civ94a} O. Civitarese and J. Suhonen , J. Phys. G: Nucl. Part.
Phys. {\bf 20} (1994) 1441.
\bibitem{Civ94b} O. Civitarese and J. Suhonen , Nucl. Phys. {\bf A 578}
(1994) 62.
\bibitem{Civ95} O. Civitarese, J. Suhonen and Amand Faessler, Nucl.
Phys. {\bf A 591} (1995) 195.
\bibitem{Par65} J. C. Parikh, Nucl. Phys. {\bf 63} (1965) 214.
\bibitem{Hec65} K. T. Hecht, Nucl. Phys. {\bf 63} (1965) 177.
\bibitem{Kle91} A. Klein and E. R. Marshalek, Rev. Mod. Phys. {\bf 63}
(1991) 375.
\bibitem{Kis63} L. S. Kisslinger and R. A. Sorensen, Rev. Mod.
Phys. {\bf 35} (1963) 853.
\bibitem{Aga68} D. Agassi, Nucl. Phys {\bf A 116} (1968) 49.
\bibitem{Sch68} D. Sch\"utte and K. Bleuler, Nucl. Phys. {\bf A
119} (1968) 221.
\bibitem{Row70} D. J. Rowe, {\em Nuclear Collective Motion},
Methuen and Co. Ltd., London 1970.
\bibitem{Mar90} A. Mariano, J. Hirsch and F. Krmpoti\'c, Nucl. Phys.
{\bf A 518} (1990) 523 and references therein.
\bibitem{Eng96} J. Engel, S. Pittel, M. Stoitsov, P. Vogel and J. Dukelsky,
Los Alamos Preprint nucl-th/9610045.
\bibitem{Duk96} J. Dukelsky, P. Schuck, Phys. Lett. {\bf B 387} (1996) 233.
\end{thebibliography}
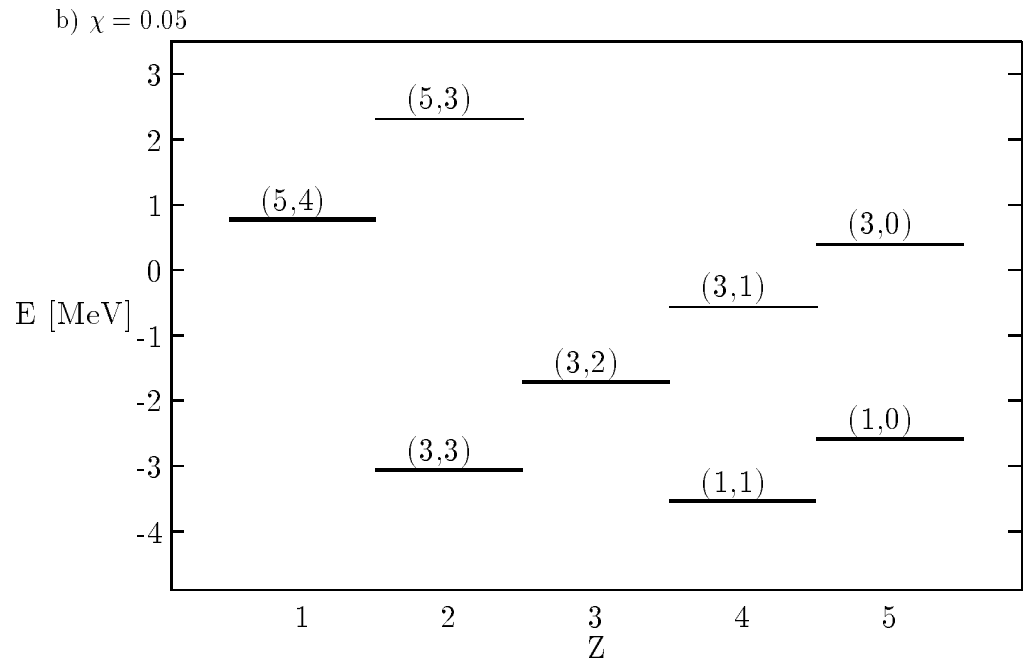
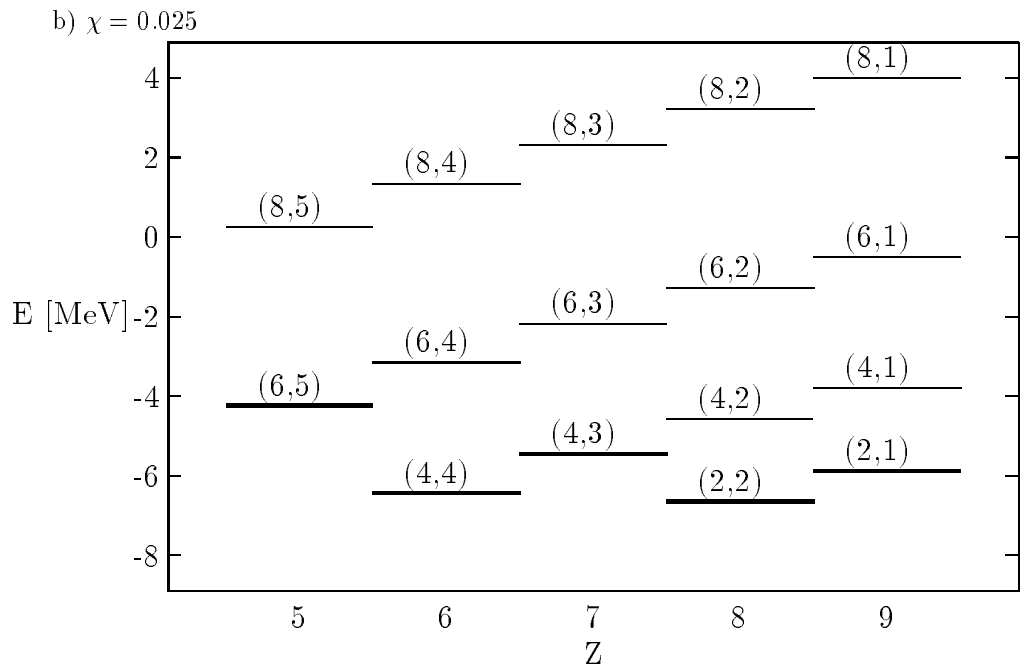
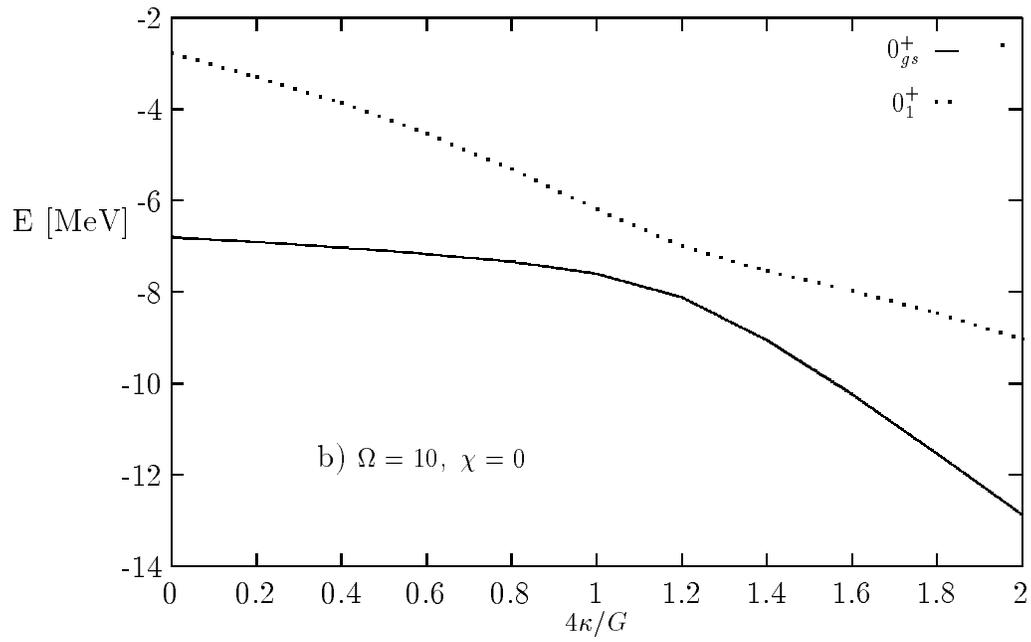
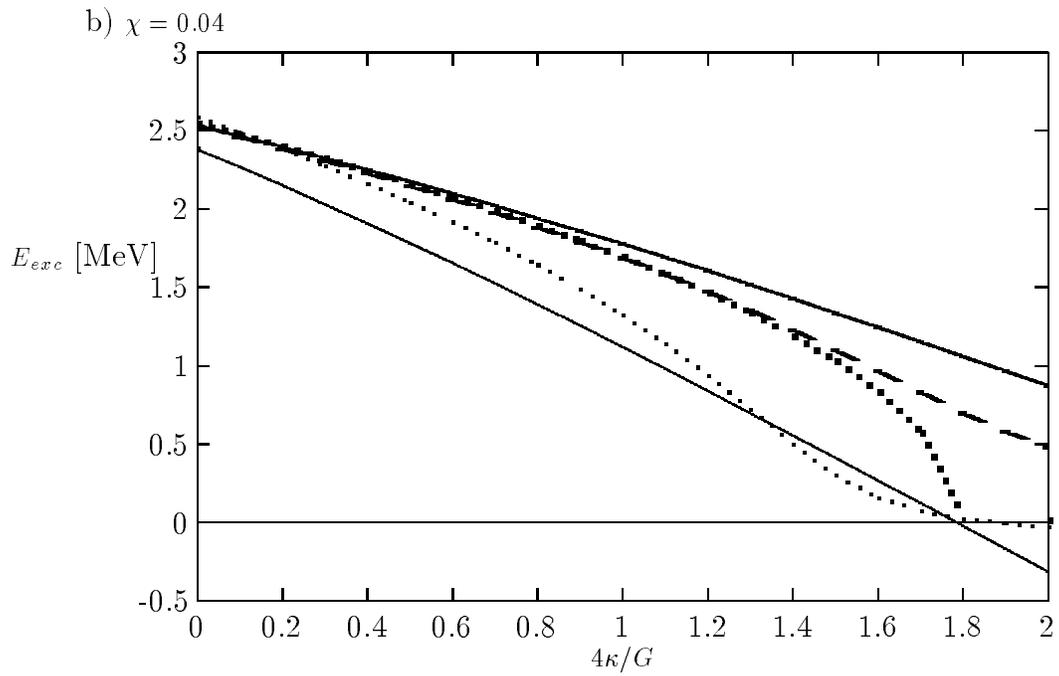
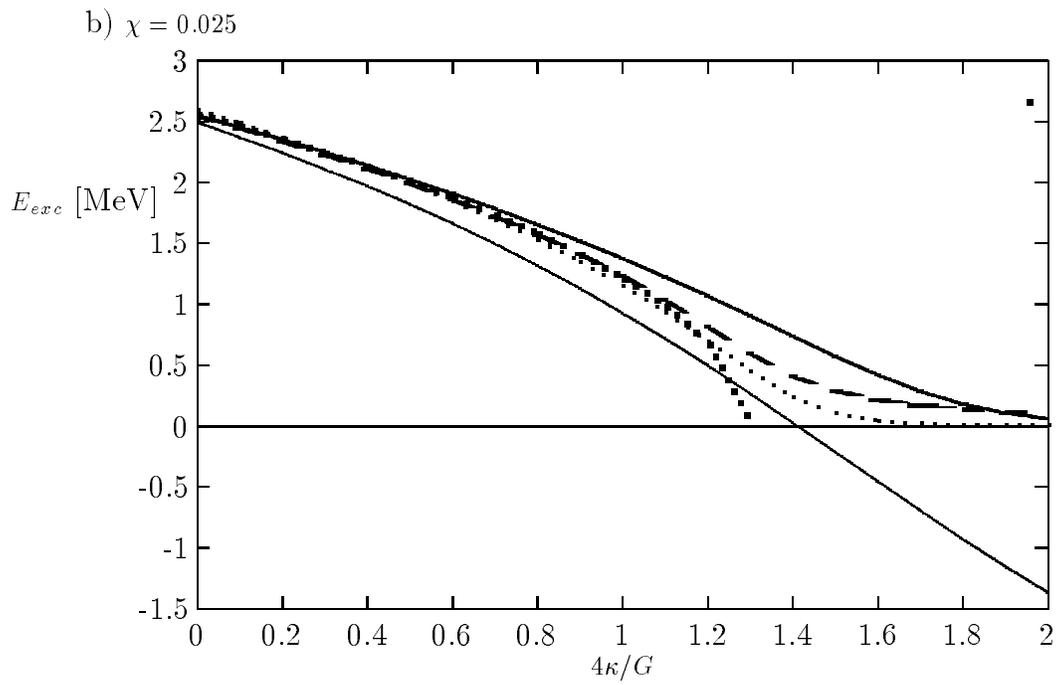
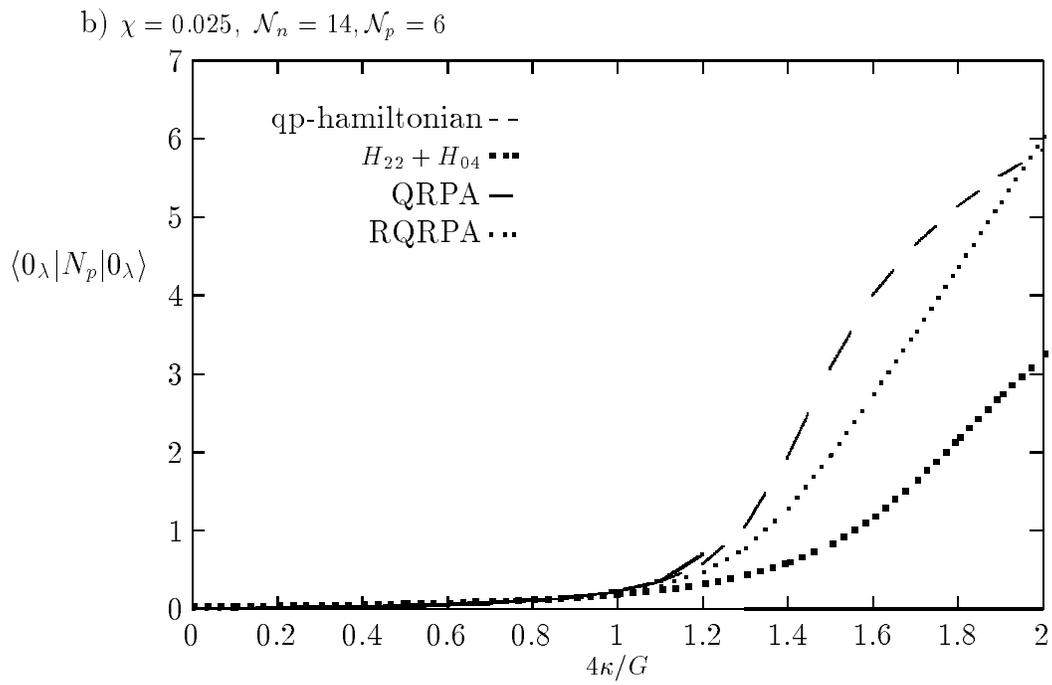
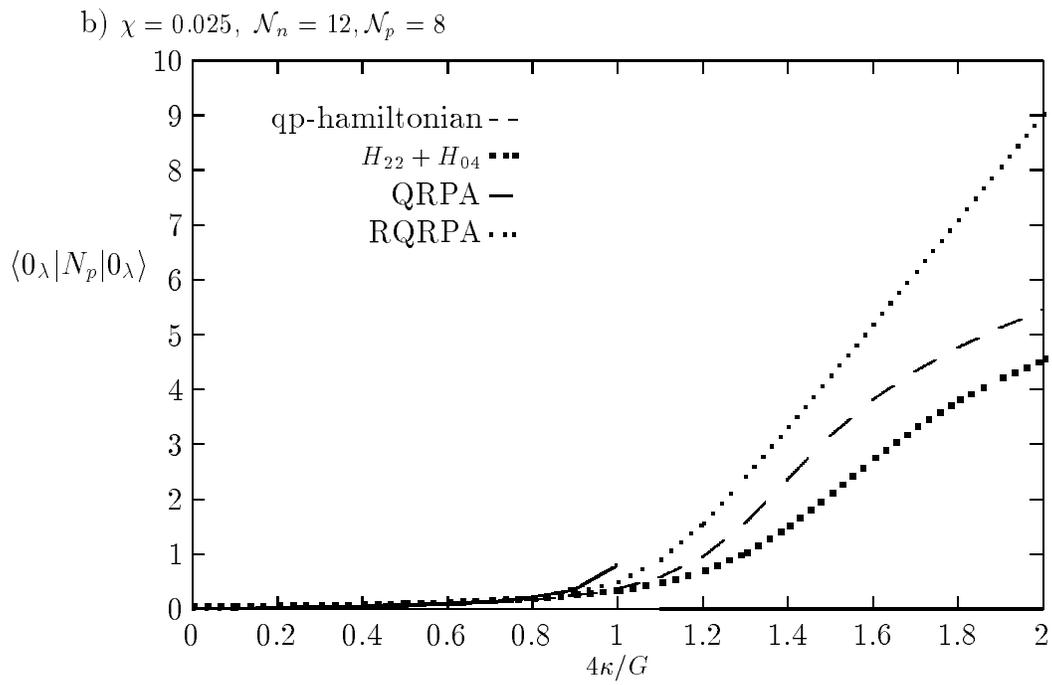
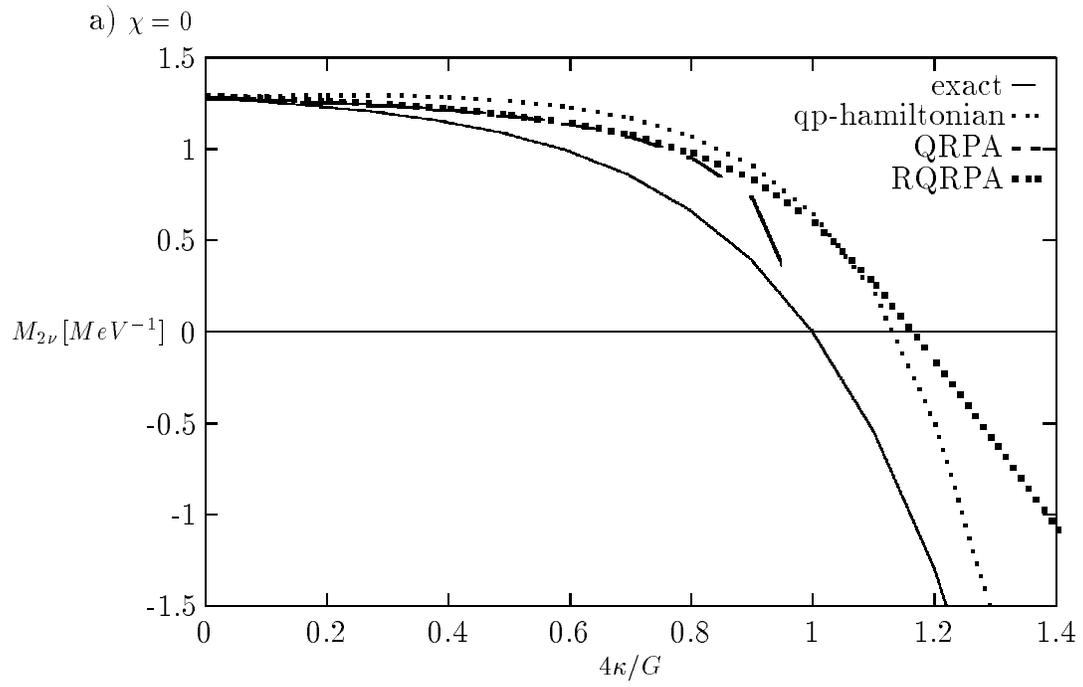
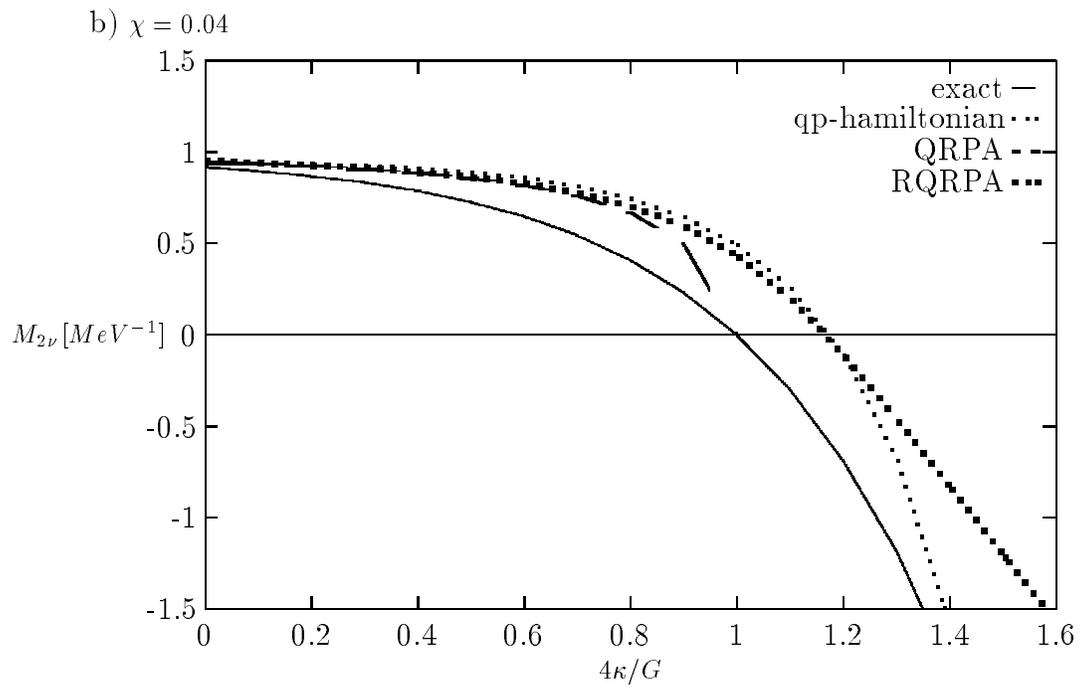
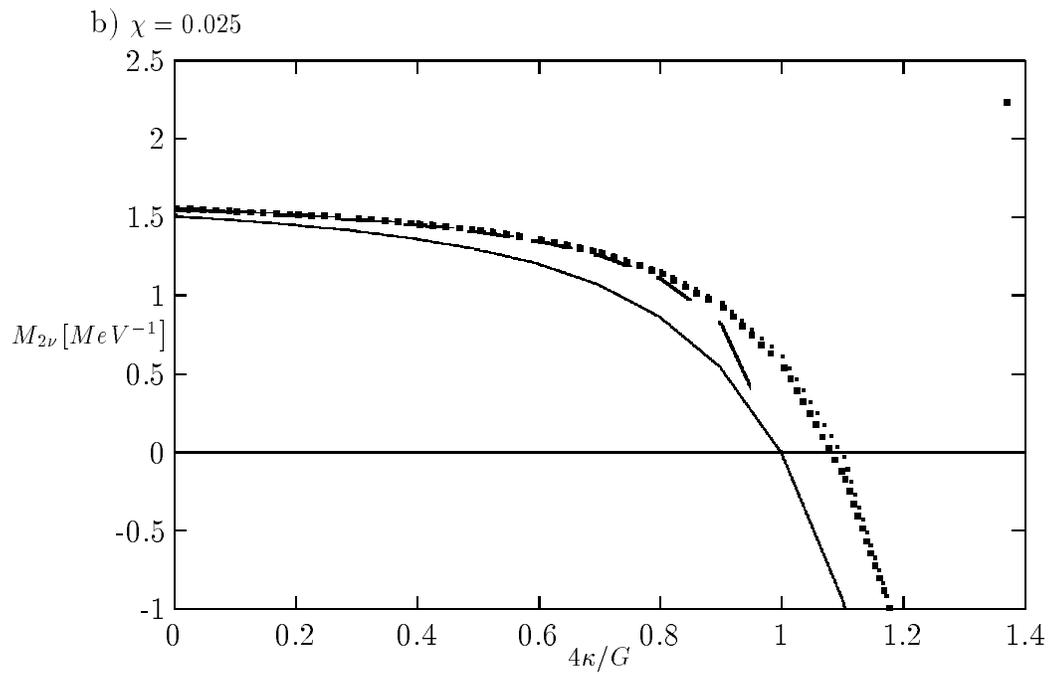
\end{document}